\documentclass[a4paper,11pt]{article}
\pdfoutput=1

\usepackage[sumlimits,intlimits,namelimits]{amsmath}
\usepackage{amssymb}
\usepackage[english]{babel}
\usepackage{bbm}
\usepackage{parskip}

\usepackage{graphicx}          %erlaubt Graphiken einzubinden

\usepackage{cite}

\usepackage[makeroom]{cancel}

\usepackage[usenames,dvipsnames]{color}

 \usepackage[bordercolor=black,backgroundcolor=yellow,linecolor=black,
textsize=scriptsize,shadow]{todonotes}

\newcommand{\AdS}{$\text{AdS}_{3}$ }

\newcommand{\arctanh}{\text{arctanh}}
\newcommand{\arcsinh}{\text{arcsinh}}

\newcommand{\mS}{\mathcal{S}}
\newcommand{\mL}{\mathcal{L}}

\newcommand{\mK}{\mathcal{K}}
\newcommand{\mN}{\mathcal{N}}

\newcommand{\totd}{\text{d}}

\newcommand{\Poincare}{Poincar\'e }

\global\parskip 6pt
\newcommand{\qed}{\nobreak \ifvmode \relax \else\ifdim\lastskip<1.5em 
\hskip-\lastskip\hskip1.5em plus0em minus0.5em \fi \nobreak\vrule height0.75em 
width0.5em depth0.25em\fi}

\newcommand{\be}{\begin{eqnarray}}
\newcommand{\ee}{\end{eqnarray}}

\def\>{\rangle}
\def\<{\langle}

\newcommand{\executeiffilenewer}[3]{%
\ifnum\pdfstrcmp{\pdffilemoddate{#1}}%
{\pdffilemoddate{#2}}>0%
{\immediate\write18{#3}}\fi%
}

\graphicspath{{pics/}}

\usepackage{jheppub} % for details on the use of the package, please
\usepackage[T1]{fontenc} % if needed

\title{\boldmath Bending branes for DCFT in two dimensions}
\preprint{MPP-2014-372}

 \author{Johanna Erdmenger,}
 \author{Mario Flory}
  \author{and Max-Niklas Newrzella}
 \affiliation{Max-Planck-Institut f\"ur Physik 
(Werner-Heisenberg-Institut),\\F\"ohringer Ring 6, D-80805, Munich, Germany}

\emailAdd{jke@mpp.mpg.de}
\emailAdd{mflory@mpp.mpg.de}
\emailAdd{maxnew@mpp.mpg.de}

\abstract{We consider a holographic dual model for defect conformal field 
theories (DCFT) in which we include the backreaction of the defect on the dual 
geometry. In particular, we consider a dual gravity system in which a 
two-dimensional hypersurface with matter fields, the brane, is embedded into a 
three-dimensional asymptotically Anti-de Sitter spacetime. Motivated by recent 
proposals for holographic duals of boundary conformal field theories (BCFT), 
we assume the geometry of the brane to be determined by Israel junction 
conditions. We show that these conditions are intimately related to the energy 
conditions for the brane matter fields, and explain how these energy 
conditions constrain the possible geometries. This has implications for the 
holographic entanglement entropy in particular. Moreover, we give exact 
analytical solutions for the case where the matter content of the brane is a 
perfect fluid, which in a particular case corresponds to a free massless 
scalar field. Finally, we describe how our results may be particularly useful 
for extending a recent proposal for a holographic Kondo model. }

\begin{document} 
\maketitle
\flushbottom

%%%%%%%%%%%%%%%%%%%%%%%%%%%%%%%%%%%%%%%%%%%%%%%%%%%%%%%%%%%%%%%%%%%%%%%%%%%%%%%%
%%%%%%%%%%%%%%%%%%%%%%%%%%%%%%%%%%%%%%%%%%%%%%%%%%%%%%%%%%%%%%%%%%%%%%%%%%%%%%%%

\section{Introduction}
\label{sec::intro}

There are important examples of physical systems which contain degrees of
freedom confined to a lower-dimensional subspace of spacetime and
can be described by conformal field theories.
This was studied in particular for CFT in 1+1 dimensions, where often the 
defect may be described in terms of a boundary condition \cite{Affleck:1995ge}.
In the case where the ambient degrees of freedom are strongly 
coupled even without the defect, it is natural to expect that applying the
AdS/CFT correspondence \cite{Maldacena:1997re,Gubser:1998bc,Witten:1998qj} or 
one of its extensions will also prove very useful in this context.

The most general holographic approach for studying defect conformal
field theory (DCFT) is to consider `Janus' solutions in supergravity, which may 
be embedded in string theory \cite{Aharony:2003qf,Bak:2003jk,D'Hoker:2007xy} 
(for recent work see
\cite{Chiodaroli:2012vc,Chiodaroli:2011fn,Gutperle:2012hy,Jensen:2013lxa,
Dias:2013bwa,Korovin:2013gha,Estes:2014hka,Azeyanagi:2007qj,
Bak:2011ga,Bak:2007jm}). The Janus configurations amount to a
$d$-dimensional domain wall embedded in AdS$_{d+1}$,  for which the
field theory defect dissolves smoothly in the interior of the spacetime.
Due to the non-local structure of these configurations, the calculation of
physical observables is involved in this approach and generally
involves PDE's.

Recently, there has been interest in the holographic study of boundary 
conformal field theories (BCFT) and defect conformal field theories
(DCFT) in a simpler approach where the gravity dual of the boundary or
defect is taken to be a localised infinitesimally thin $d$-dimensional surface 
in AdS$_{d+1}$ \cite{Takayanagi:2011zk,Fujita:2011fp,Nozaki:2012qd}. These 
models were generalised to holographic models of boundary RG flows, where 
`boundary' refers to the field theory boundary degrees of freedom rather than to 
the AdS boundary \cite{Takayanagi:2011zk,Fujita:2011fp,Nakayama:2012ed}.
It was shown that these flows satisfy a holographic version of the
g-theorem, which states that the boundary entropy decreases along an
RG flow from a UV to an IR fixed point
\cite{Takayanagi:2011zk,Fujita:2011fp,Nozaki:2012qd,Nakayama:2012ed}.
Within field theory, this theorem was given in
\cite{Affleck:1991tk,Friedan:2003yc}.

Gravity duals of DCFT in which the dual of the defect remains
localised were recently used also for holographic models of the
quantum Hall effect. Recent examples include both bottom-up models
\cite{Melnikov:2012tb,Lippert:2014jma} and top-down models involving brane 
constructions \cite{Kristjansen:2013hma,Bergman:2010gm}. For example, these 
models display incompressible states with quantised Hall conductivity.

In this paper, we find explicit solutions of the models of 
\cite{Takayanagi:2011zk,Fujita:2011fp,Nozaki:2012qd} for non-trivial field 
content. We refer to the localised gravity dual of the field theory 
defect as the {\it brane}, and find explicit solutions which take the 
backreaction of the brane matter fields on the geometry into account. We 
construct holographic duals for DCFT by gluing 
together two a priori distinct asymptotically AdS manifolds. 
For this to be consistent with the Einstein field 
equations, the system has  to obey the Israel junction 
conditions \cite{Israel:1966rt}. These relate the exterior curvature and the 
induced metric of the brane with the energy-momentum tensor 
constrained to it. The defect energy-momentum tensor is dynamically 
generated by matter fields on the brane, subject to appropriate 
energy conditions. Throughout the paper, we will always assume the null 
energy condition (NEC) to be satisfied and also investigate the impact of the 
strong (SEC) and weak energy condition (WEC).

In the AdS/BCFT models mentioned above \cite{Takayanagi:2011zk,
Fujita:2011fp,Nozaki:2012qd},  von Neumann boundary conditions are imposed on 
the brane and, as usual, Dirichlet conditions on the conformal boundary. This 
ansatz yields the same equations of motion as the use of Israel junction 
conditions together with the assumption that there is a reflection symmetry 
around the brane. This is analogous to classical electrodynamics, where 
von Neumann boundary conditions may be imposed by introducing mirror charges.

Moreover, in this paper  we use the Israel junction conditions to link energy 
conditions imposed on the brane matter to qualitative statements about 
the exterior geometry of its embedding into the ambient spacetime.
In particular, we make use of the so-called `barrier theorem' recently 
proved by Engelhardt and Wall in \cite{Engelhardt:2013tra}. This theorem states 
under which conditions the brane may be intersected by spacelike hypersurfaces 
which are anchored at the boundary of the manifold. We find a connection between 
the assumption made for this theorem and a specific combination of energy 
conditions, which can also be used to distinguish between hypersurface 
configurations which are anchored twice at the boundary, and others which are 
infinitely extended into the bulk.

We obtain analytic solutions for the embedding functions for the cases of
a constant brane tension, as well as for perfect fluids and a free massless 
scalar field in \AdS and BTZ backgrounds. We find that the constant tension 
solutions may be generated by following a normal flow starting from the trivial 
solution in a non-backreacting geometry. 

Our main motivation to study the models presented, with bulk spacetimes to 
both sides of the brane, is the holographic bottom-up model for the Kondo effect 
proposed in \cite{Erdmenger:2013dpa}, in which a magnetic impurity with $SU(N)$ 
spin interacts with a strongly coupled field theory. This model is based on a 
1+1-dimensional brane with matter fields that extends radially  from the AdS 
boundary into the bulk, in a BTZ black hole spacetime. In the holographic
model of \cite{Erdmenger:2013dpa}, the formation of the Kondo screening cloud 
corresponds to the formation of a condensate involving an electron and a slave 
fermion, as realised in field theory for large $N$ Kondo models for instance in 
\cite{Coleman:1986dva,PhysRevB.35.5072,PhysRevLett.90.216403,PhysRevB.69.035111}
. Moreover, this model involves the gravity dual of a boundary RG flow
triggered by the gravity dual of a marginally relevant field theory operator.

The model of \cite{Erdmenger:2013dpa} does not include the
backreaction of the brane degrees of freedom on the geometry. However,
to calculate the Ryu-Takayanagi holographic entanglement entropy (HEE)
\cite{Ryu:2006bv,Ryu:2006ef} for this model, it is necessary to include the 
backreaction: The HEE is obtained from a minimal surface which encodes 
information about the metric of the background geometry. This metric is not 
altered in the probe limit. According to the result of this present paper, 
the Israel junction conditions are then the natural choice to describe the 
backreaction that the energy-momentum localised on the brane exerts on the 
overall geometry. This allows for the calculation of the
HEE in AdS/BCFT, as was already done in \cite{Azeyanagi:2007qj}, but also in
AdS/DCFT models such as in \cite{Erdmenger:2013dpa}. Due to the generality of 
these junction conditions and of the AdS/BCFT ansatz, we expect that the 
findings of this present paper will have a much more wide range of applications 
than the holographic Kondo model mentioned above. We note that complementary 
approaches for calculating the HEE for DCFT were presented in 
\cite{Jensen:2013lxa,Estes:2014hka}, and for probe branes in 
\cite{Karch:2014ufa,Chang:2014oia}.

The outline of this paper is as follows: We begin by reviewing the AdS/BCFT 
approach in section \ref{sec::AdSBCFTreview}.  In section 
\ref{sec::backreaction} we specify our ansatz and its equations of motion. 
In the three-dimensional static case, this ansatz will generally only 
involve ODEs. Then, in section \ref{sec::ECs}, we  investigate these equations 
in detail, formulating them in a simpler way and analysing the implications of 
the energy conditions imposed on the brane. These energy conditions play an 
important role in section \ref{sec::Wall}, where we make the connection between 
our ansatz and the `barrier theorem' recently proven in 
\cite{Engelhardt:2013tra}. In section \ref{sec::toymodel} we  revisit a toy 
model with constant brane tension previously studied in \cite{Azeyanagi:2007qj} 
and reformulate it in terms of the Israel junction conditions. We then apply our 
methods to the case where the matter content on the brane can be described by a 
perfect fluid or a free massless scalar field in \ref{sec::PF}. Moreover, in 
section \ref{sec::Kondo} we study a more involved matter content which behaves 
very differently from the fluid solutions investigated before, and is physically 
motivated as a holographic dual of the Kondo effect as introduced in 
\cite{Erdmenger:2013dpa}. We conclude with an outlook, in particular on 
calculating the HEE for the holographic Kondo model of 
\cite{Erdmenger:2013dpa}, in section \ref{sec::conc}. We comment on the 
generalisation of the results from section \ref{sec::Wall} to higher dimensions 
in appendix \ref{sec::Wall2}. Finally, in appendix \ref{app::normalflow} we give 
precise statements about if and under which assumptions the construction of 
constant brane tension solutions can be generalised to other spacetimes.

%%%%%%%%%%%%%%%%%%%%%%%%%%%%%%%%%%%%%%%%%%%%%%%%%%%%%%%%%%%%%%%%%%%%%%%%%%%%%%%%
%%%%%%%%%%%%%%%%%%%%%%%%%%%%%%%%%%%%%%%%%%%%%%%%%%%%%%%%%%%%%%%%%%%%%%%%%%%%%%%%

\section{Review of the AdS/BCFT formalism}
\label{sec::AdSBCFTreview}

\begin{figure}[htb]
\centering
 \def\svgwidth{0.35\columnwidth}
\executeiffilenewer{NMQP3.svg}{NMQP3.pdf}%
{inkscape -z -D --file=NMQP3.svg %
--export-pdf=NMQP3.pdf --export-latex}%
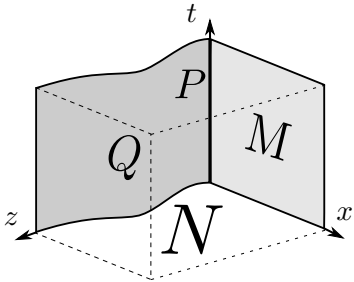%

\caption{Setup for the holographic description of BCFT: Asymptotically 
AdS bulk spacetime $N$ with conformal boundary $M$ and additional boundary $Q$. 
$P$ is the intersection of $M$ and $Q$. On the field theory 
side, we refer to $P$ as the defect and to $M$ as the ambient space. 
}
\label{fig::NMQP}
\end{figure}

We begin by reviewing the holographic approach to boundary conformal field 
theories (BCFT) proposed in 
\cite{Takayanagi:2011zk,Fujita:2011fp,Nozaki:2012qd}\footnote{This method was 
later utilised in 
\cite{Alishahiha:2011rg,Setare:2011ey,Kwon:2012tp,Fujita:2012fp,Nakayama:2012ed,
Melnikov:2012tb,Astaneh:2014fga,Magan:2014dwa}.}. 
In the standard AdS/CFT correspondence, an asymptotically AdS spacetime $N$ 
is considered, with a conformal boundary $M$ and a Dirichlet boundary condition 
imposed on the bulk metric at $M$. In contrast to this, the AdS/BCFT ansatz 
mentioned above introduces an additional boundary $Q$ that intersects $M$ in $P$ 
and extends from there into the bulk as shown in figure \ref{fig::NMQP}. On this 
hypersurface $Q$ (which we will also refer to as brane), the bulk-metric will be 
required to satisfy a von Neumann 
boundary condition.

Let us consider the metric contribution to the dual gravity action. With the 
usual Gibbons-Hawking boundary term \cite{Gibbons:1976ue} this action reads 
\cite{Takayanagi:2011zk,Fujita:2011fp,Nozaki:2012qd}
\begin{equation}
\begin{aligned}
 \mS = \,&\frac{1}{2\kappa}\int_N \totd^{d+1}x\, 
    \sqrt{-g}\left(R-2\Lambda\right)
    +\frac{1}{\kappa}\int_M \totd^d x\, \sqrt{-h} 
    \left(K^{(h)}-\Sigma^{(h)}\right)\\  
 &+\frac{1}{\kappa}\int_Q \totd^d x\, \sqrt{-\gamma} 
    K^{(\gamma)}
 +\int_Q\totd^d x\,\sqrt{-\gamma} \,\mL_Q
    + S_{P},
\end{aligned}
\label{Takayanagi}
\end{equation}
where we defined $\kappa = 8 \pi G_N$. The Lagrangian $\mL_Q$
describes matter fields constrained to $Q$, and $S_{P}$ the necessary 
counterterms arising on $P$. Any extrinsic curvature is defined
via normal vectors pointing outwards. 

In \eqref{Takayanagi} the first line contains the usual metric bulk- and 
boundary-terms from AdS/CFT correspondence, with $K^{(h)}$ the extrinsic 
curvature of the induced metric $h_{ij}$ on $M$. The constant $\Sigma^{(h)}$ is 
added as a counterterm for holographic renormalisation. The new terms in the 
AdS/BCFT ansatz are those in the second line of \eqref{Takayanagi}: The first 
two describe the extrinsic geometry of the brane $Q$ and the dynamics of matter 
fields with Lagrangian $\mL_{Q}$ possibly living on it, while $S_P$ describes 
boundary terms that may arise from $P$. We will explicitly allow for a 
cosmological constant (or constant tension) term $\mL_{Q}=const.$, as this is 
one of the most widely studied models of AdS/BCFT 
\cite{Takayanagi:2011zk,Fujita:2011fp,Nozaki:2012qd}. 
Taking the variation of this action, with Dirichlet boundary conditions on $M$ 
and von Neumann boundary conditions on $Q$ yields the usual bulk equations of 
motion together with the boundary condition
\begin{align}
 K^{(\gamma)}_{ij}-\gamma_{ij}K^{(\gamma)}=\kappa S_{ij},
 \label{Takayanagi2}
\end{align}
with $S_{ij}$ the energy-momentum tensor derived from $\mL_{Q}$, 
the letter $S$ referring to `shell' or `surface'. 

As starting point for the present paper, we note that while the 
equations \eqref{Takayanagi2} are derived as boundary conditions from the 
surface term $\int_Q ...$, they take a form similar to the Israel junction 
conditions \cite{Israel:1966rt} that describe the matching of 
\textit{two} spacetimes along the hypersurface $Q$. These conditions will be 
described in section \ref{sec::backreaction}. In this setting, the 
hypersurface $Q$ has a specified embedding in both spacetimes, with 
corresponding points on $Q$ being identified, see figure \ref{fig::geometry} 
for an illustration. An observer who lives in one of the spacetimes and enters 
$Q$ will hence emerge in the other spacetime. As shown by Israel 
\cite{Israel:1966rt}, such a sewing of two spacetimes along a given 
hypersurface $Q$ can only be sustained by Einstein's equations if there 
is a correct distribution of energy-momentum localised on $Q$. Under a certain 
symmetry assumption, these Israel junction conditions will then take a form 
similar to \eqref{Takayanagi2}, as we will show in more detail in section 
\ref{sec::backreaction}. We will propose a holographic model for defect CFT 
(DCFT) which follows this line of thought.\footnote{Indeed, during the 
preparation of this work similar ideas have been 
proposed in \cite{Magan:2014dwa}.
}

While the setup \eqref{Takayanagi} has been studied in a number 
of papers, concrete analytical solutions for non-constant $\mL_{matter,\,Q}$ 
are quite rare in the literature, see \cite{Magan:2014dwa} for one recent 
exception. Hence the explicit solutions presented in sections \ref{sec::PF} 
and \ref{sec::Kondo} below, as well as the methods presented in section 
\ref{sec::ECs} that allow us to obtain these solutions quite simply, are the 
main results of this paper.

%%%%%%%%%%%%%%%%%%%%%%%%%%%%%%%%%%%%%%%%%%%%%%%%%%%%%%%%%%%%%%%%%%%%%%%%%%%%%%%%
%%%%%%%%%%%%%%%%%%%%%%%%%%%%%%%%%%%%%%%%%%%%%%%%%%%%%%%%%%%%%%%%%%%%%%%%%%%%%%%%

\section{A model for DCFT in two dimensions}
\label{sec::backreaction}

\begin{figure}[ht!]
 \centering
 \def\svgwidth{0.8\columnwidth}
\executeiffilenewer{geometryNEW3.svg}{geometryNEW3.pdf}%
{inkscape -z -D --file=geometryNEW3.svg %
--export-pdf=geometryNEW3.pdf --export-latex}%
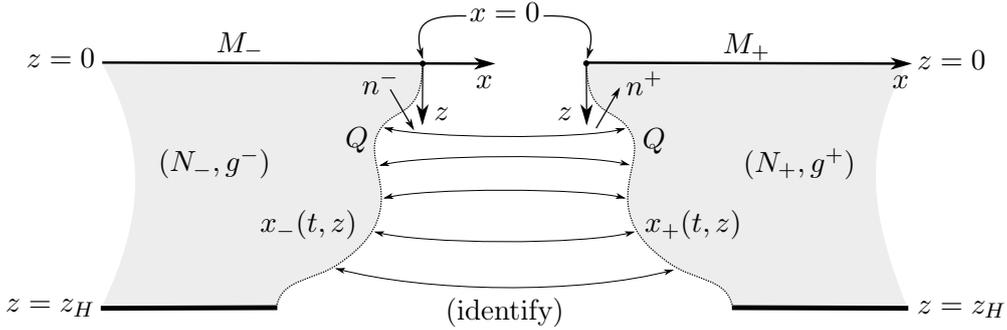%

\caption{Geometry of the setup: The manifold $N$ is split into two 
submanifolds $N_+$ and $N_-$. The white region is excised from the
manifold. For each submanifold, the position of the brane is given by 
$x=x_{\pm}(t,z)$ and corresponding points of the two embeddings $x_+$ and $x_-$ 
are identified, as indicated by double arrows. The normal vectors to $Q$ on 
both sides are named $n^{\pm}$, and point from $N_-$ to $N_+$. In 
most of the paper, we will assume $N_{\pm}$ to be BTZ metrics \eqref{BTZmetric} 
of equal temperature and the embedding to be symmetric, i.e.~$x_+=-x_-$.}
\label{fig::geometry}
\end{figure}

In this section we describe in detail the setup we are going to use in 
this paper. The idea is to take the AdS/BCFT setup depicted in figure 
\ref{fig::NMQP} (with an additional boundary $Q$ that reaches from the AdS 
boundary $M$ into the bulk) and to allow for bulk spacetimes $N_\pm$ on both 
sides of $Q$ as in figure \ref{fig::geometry}. 

Indeed, the equations \eqref{Takayanagi2} resemble the \textit{Israel junction 
conditions} that describe how two spacetimes can be sewn together. For example, 
starting somewhere in $N_+$ in figure \ref{fig::geometry}, a curve might cross 
$Q$ and leave it immediately into $N_-$ to the left, as the corresponding points 
of $Q$ in $N_+$ and $N_-$ are identified. In this setup, the hypersurface $Q$ is 
not a boundary of the bulk spacetime, but merely a hypersurface (or brane) 
embedded into a spacetime that consists of the two halves $N_\pm$.

What is the physical interpretation of such a setup? We know
that in $2+1$-bulk dimensions, Einstein-Hilbert gravity does not have any 
propagating bulk degrees of freedom and all vacuum solutions (including the 
famous BTZ black holes \cite{Banados:1992wn,Banados:1992gq}) will hence be 
locally identical to the \AdS metric. Suppose now we have a holographic 
(toy) model of a defect CFT (DCFT) that involves a hypersurface that is anchored 
at the boundary and reaches into the bulk, such as the one proposed in 
\cite{Erdmenger:2013dpa}. If there are matter fields living on this 
hypersurface, how will their energy-momentum backreact on the 
bulk spacetime? Here we investigate this question in the case of three 
bulk dimensions, where gravity has no propagating degrees of freedom and the 
bulk spacetime has to be locally $\text{AdS}_3$.

Our ansatz is that the backreaction of such a 
model will be given precisely by the Israel junction conditions. The right 
hand side of the brane $Q$ will be the locally \AdS spacetime $N_+$, the left 
hand side of the brane will be the spacetime $N_-$. We define the representation
of $Q$ in $N^{\pm}$ by two a priori distinct embeddings
\begin{equation}
 X^{\pm}: Q\hookrightarrow N^{\pm},\ \  
 X^{\pm}(t,z) = \left( t , z , x^{\pm}(t,z)) \right),
\end{equation}
which we assume to be differentiable at least once.

The entire spacetime is then constructed by identifying corresponding points 
along the curves $x_+$ and $x_-$, see figure \ref{fig::geometry}.\footnote{Note 
that in the construction presented here, the induced metrics on both embeddings 
of $Q$ are assumed to be equal: $\gamma_-=\gamma_+\equiv \gamma$. Hence 
choosing a specific coordinate system on $Q$, two points in the two embeddings 
of $Q$ which are described by the same coordinates on $Q$ are to be identified. 
See \cite{Marolf:2005sr} for an exploration of the possibility 
$\gamma_+\neq\gamma_-$.} Assuming that no external force acts on the brane, the 
Israel junction conditions then relate the functions $x_\pm(t,z)$ with the 
energy-momentum tensor $S_{ij}$ on the brane \cite{Israel:1966rt,Battye:2001pb},
\begin{align}
 \left\{K^{ij}\right\}S_{ij}&=0,
 \label{eq:Israeljunctionconditions2}\\
 \left[K_{ij}-\gamma_{ij}\,\text{tr}\,K\right]&=-\kappa\,S_{ij} ,
 \label{eq:Israeljunctionconditions}
\end{align}
where $\{A\,\}:=\frac{1}{2}(A^+ + A^-)$, $[A\,]:=A^+ - A^-$ for any tensors 
$A^{\pm}$ defined on both sides of the brane. $\gamma_{ij}$ is the induced 
metric on the brane, $S_{ij}$ is the energy-momentum tensor on the brane and 
$\kappa$ the coupling constant in the Einstein equations. 

Given the embeddings 
$x_\pm$, we may define two functions $f^{\pm}(t,z,x) := x - x_{\pm}(t,z)$ in 
$N_{\pm}$ and obtain two one-forms by applying the exterior derivative $\mN 
\mathfrak{n}^{\pm} = \totd f^{\pm}$ with $\mN$ a normalisation factor. The 
corresponding vectors defined via $\mathfrak{n}^{\pm}(n^{\pm})=1$ give the 
normals to the embedding. Due to our sign convention in the definition 
of $f^{\pm}$, these normal vectors point out of $N^-$ and into $N^+$, see 
figure \ref{fig::geometry}.
Using the definition of the extrinsic curvature 
\begin{equation}
K^{\pm}(u,v) = g\left(\nabla^{\pm}_{U^{\pm}} V^{\pm},n^{\pm}\right)
=-g\left(V^{\pm},\nabla^{\pm}_{U^{\pm}} n^{\pm}\right),
\end{equation}
where $U^{\pm}$ and $V^{\pm}$ denote the pushforwards of $u$ and $v$ 
by $X^{\pm}$, we compute its components in this specific coordinate system. 
In components, the definition is given by
\begin{equation}
 K^{\pm}_{ij}\equiv \frac{\partial X^{\alpha}}{\partial \xi^i} 
\frac{\partial X^{\beta}}{\partial \xi^j} \nabla_{\alpha} 
 \mathfrak{n}^{\pm}_{\beta} = -\mathfrak{n}_{\alpha}^{\pm}
 \left( \frac{\partial^2 X^{\alpha}}{\partial \xi^i \partial 
\xi^j} 
 + \Gamma_{\beta\gamma}^{\alpha}\frac{\partial X^{\beta}}{\partial \xi^i} 
\frac{\partial X^{\gamma}}{\partial \xi^j} \right),
\label{ExtrinsicCurvature}
\end{equation}
where $\xi^{i}$ denote the coordinates on $Q$ and 
$\Gamma^{\alpha}_{\beta\gamma}$ the Christoffel symbols in $N^{\pm}$.

In view of applications to systems with finite temperature, we 
assume the bulk spacetimes $N^{\pm}$ to be BTZ black holes equal temperature, 
i.e.~the metrics $g^{\pm}$ are identical and given by 
\begin{align}
 g^{\pm}_{\mu\nu}\totd x^\mu \totd x^\nu =: g_{\mu\nu}\totd x^\mu \totd x^\nu
 =\frac{L^2}{z^2}\left(-h(z) \totd t^2 + \frac{\totd z^2}{h(z)}+\totd 
x^2\right),
\label{BTZmetric}
\end{align}
with $h(z)=1-z^2/z^2_H$ for the BTZ black hole 
and $h(z)=1$ in the limiting \Poincare \AdS  case. With this, and choosing the 
coordinates on $Q$ to be $t,z$, we explicitly find
\begin{equation}
 \begin{aligned}
 K^\pm_{tt}&=\frac{1}{\mathcal{N}}\left(
  \partial_t^{\,2}  x_{\pm} -\partial_z x_{\pm}\left( 
\frac{h}{2}\left(h'-\frac{2h}{z}\right)+\frac{h}{z} (\partial_t x_{\pm})^2 
\right)\right),\\
 K^\pm_{tz}=K_{zt}&=\frac{1}{\mathcal{N}}\left(
  \partial_t \partial_z x_{\pm}- 
  \partial_t x_{\pm}\left( \frac{h}{z}(\partial_z x_{\pm})^2 + \frac{h'}{2h} 
\right)\right),\\
 K^\pm_{zz}&=\frac{1}{\mathcal{N}}\left(
  \partial_z^{\,2} x_{\pm}- 
  \partial_z x_{\pm}\left(
  \frac{h}{z} (\partial_z x_{\pm})^2
  +\frac{1}{z}
  - \frac{h'}{2h}
  \right)\right),
 \end{aligned}
\label{eq:exteriorcurvature}
\end{equation}
where $(...)'$ denotes the derivative with respect to $z$ and $\mN$ is the 
normalisation of $n^{\mu}$. The induced metric reads
\begin{equation}
 \gamma_{ij}=
 \begin{pmatrix}
  g_{tt}+\left(\partial_t x_{\pm}\right)^2 g_{xx} 				
	
& \left(\partial_t x_{\pm}\right)\left(\partial_z x_{\pm}\right) g_{xx} \\
  \left(\partial_t x_{\pm}\right)\left(\partial_z x_{\pm}\right) g_{xx} 	
	
& g_{zz}+\left(\partial_z x_{\pm}\right)^2 g_{xx}
 \end{pmatrix}.
 \label{eq:pullbackMetric}
\end{equation}
For simplicity, in the following we will always assume a 
symmetric embedding with $x_+=-x_-$, and hence find
\begin{equation}
 K^{-}_{ij} = -K^+_{ij} , \qquad \{ K \} = 0.
 \label{eq:extrinsicBrackets}
\end{equation}
For this ansatz, \eqref{eq:Israeljunctionconditions2} is 
trivially satisfied while \eqref{eq:Israeljunctionconditions} reduces to 
\begin{equation}
 K^{+}_{ij} - \gamma_{ij} K^+ = 
-\frac{\kappa}{2}\,S_{ij}
 \label{eq:leftoverIsrael}.
\end{equation}
Apart from a factor $1/2$ (due to $K^+-K^-=2K^+$), this equation seems to 
have an additional  $-$ sign as compared to \eqref{Takayanagi2}. The reason is 
that we need to find a convention that fixes the sign of the normal vector 
$n^{\mu}$. In \eqref{eq:Israeljunctionconditions} the signs are correct when 
assuming that $n^{\mu}$ points from $N_-$ to $N_+$. Then expressing 
\eqref{eq:Israeljunctionconditions} (with $x_+=-x_-$) in 
terms of $K^+$ leads to \eqref{eq:leftoverIsrael}. The equation 
\eqref{Takayanagi2} in contrast was derived from the Gibbons-Hawking 
like boundary term in \eqref{Takayanagi}, where $n^{\mu}$ was defined to 
point out of the bulk. Hence alternatively expressing \eqref{eq:leftoverIsrael} 
in terms of $K^-$, the signs would be as in \eqref{Takayanagi2}. 

The Israel junction conditions may also be derived from the variational 
problem using the Einstein-Hilbert action for both parts of the spacetime,
including the Gibbons-Hawking terms, and identifying their common boundary $Q$.
For details see \cite{Hayward:1990tz,Hayward:1993my,Chakkrit}.

In this paper we allow for a two-sided defect setup as depicted in figure 
\ref{fig::geometry}. Due to the similarity of the equations \eqref{Takayanagi2} 
and \eqref{eq:leftoverIsrael}, our results will equally apply to the one-sided 
AdS/BCFT setup \eqref{Takayanagi2}. Our results will have applications on the 
holographic bottom-up models of $1+1$-dimensional BCFT or DCFT that contain (or 
can be approximated by bulk spacetimes containing) co-dimension one 
hypersurfaces with a non-trivial matter content, such as for example 
\cite{Erdmenger:2013dpa}. For other holographic (often top-down) studies of BCFT 
and DCFT, see 
\cite{Chiodaroli:2012vc,Chiodaroli:2011fn,Gutperle:2012hy,Jensen:2013lxa,
Dias:2013bwa,Korovin:2013gha,Estes:2014hka}\footnote{In a not 
necessarily holographic context, the Israel junction conditions may also play a 
role in brane world scenarios, see 
\cite{Randall:1999vf,Karch:2000ct,Karch:2000gx,Battye:2001pb}.}.

%%%%%%%%%%%%%%%%%%%%%%%%%%%%%%%%%%%%%%%%%%%%%%%%%%%%%%%%%%%%%%%%%%%%%%%%%%%%%%%%
%%%%%%%%%%%%%%%%%%%%%%%%%%%%%%%%%%%%%%%%%%%%%%%%%%%%%%%%%%%%%%%%%%%%%%%%%%%%%%%%

\section{Decomposition of the Israel junction conditions}
\label{sec::ECs}

In this section, our aim is to show how for a two-dimensional brane the 
tensorial Israel junction conditions \eqref{eq:leftoverIsrael} can be projected 
into scalar equations in a very simple manner. Writing the equations in this 
way will allow us to easily make a connection with energy conditions in 
subsection \ref{sec::ECs2} and with the barrier theorem of Engelhardt and Wall 
in section \ref{sec::Wall}. Moreover, this will allow us to find simple exact 
solutions to these equations for non-trivial matter content on the brane in 
section \ref{sec::PF}.

\subsection{The brane energy-momentum tensor in two dimensions}
\label{sec::decomposition}

As the brane worldsheet is 1+1-dimensional, the energy-momentum tensor 
as a symmetric $(0,2)$-tensor is described by three values at every point in the 
spacetime. In order to achieve a decomposition of the 
equation \eqref{eq:leftoverIsrael} into scalar equations, we define a 
basis of three tensors that span the space of symmetric $(0,2)$-tensors in 1+1 
dimensions. To this end, we note that the lightcone on the 
1+1-dimensional brane worldsheet does not consist of infinitely many 
null-vectors, but of two distinct null directions, which we normalise such that 
the two (left- and right pointing) null vectors satisfy
\begin{align}
 l_i l^i=0=r_i r^i,\ \ l_i r^i =-1.
\end{align}

We may now decompose the energy-momentum tensor $S_{ij}$ (and any other 
symmetric $(0,2)$-tensors on the brane) as follows,
\begin{align}
 S_{ij}=\frac{S}{2}\gamma_{ij}+S_{L}l_i l_j +S_{R} r_i r_j.
 \label{decompofS}
\end{align}
Here $\gamma_{ij}$ is the induced metric on the brane, and $S$ is the trace of 
$S_{ij}$. 
$l_i l_j$ and $r_i r_j$ are two independent symmetric traceless tensors, and 
$S_L$ 
and $S_R$ are the non-trace components of $S_{ij}$ in this 
decomposition\footnote{It is indeed easy to define proper projection operators, 
e.g.~$S_{L}=S_{ij}r^i r^j$.}.

\subsection{Energy conditions in two dimensions}
\label{sec::ECs2}

We will repeatedly make use of energy conditions on the brane energy-momentum 
tensor $S_{ij}$.  Let us discuss some of these conditions in detail in the 
following. For reviews of energy condition we refer to 
\cite{Visser:1999de,Curiel:2014zba}.

The \textit{null energy condition} (NEC) implies that for every null 
vector $m^{i}$ on the worldsheet of the brane, we have
\begin{align}
 S_{ij}m^{i}m^{j}\geq0\ \  \forall\  m^i m_i=0.
\end{align}
As said above, there are only two distinct 
null directions on the brane. The NEC hence reads 
\begin{align}
 S_{ij}l^{i}l^{j}\geq0&\text{  and  }S_{ij}r^{i}r^{j}\geq0\ \ 
%\nonumber
%\\
\Rightarrow\ \  S_L\geq0\text{  and  }S_R\geq0.
\end{align}

%\subsection{WEC}

The \textit{weak energy condition} (WEC) is similar to the NEC, just with 
timelike vectors, i.e.
\begin{align}
 S_{ij}m^{i}m^{j}\geq0\ \  \forall\  m^i m_i<0.
\end{align}
In general, parameterising $m^i=\alpha l^i + \beta r^i$ with $\alpha,\beta>0$ 
we 
find\footnote{Note that the WEC implies the NEC by 
continuity in the limits 
$\alpha\rightarrow0$ respectively $\beta\rightarrow0$.}
\begin{align}
S_L\geq0,\ \ S_R\geq0,\ \ S\alpha\beta\leq S_L \alpha^2+S_R\beta^2\ \ 
\forall \alpha,\beta>0.
\end{align}
This implies for optimal choice of $\alpha,\beta$:
\begin{align}
S_L\geq0,\ \ S_R\geq0,\ \ S\leq 2\sqrt{S_L S_R}.
\end{align}

The last energy condition that we are going to discuss is the \textit{strong 
energy condition} (SEC). Unfortunately, the proper generalisation of the SEC in 
3+1 dimensions to arbitrary dimensions is not unique, and we will find it 
most useful to set\footnote{Theories of spacetime curvature are usually 
described by equations that equate a matter energy-momentum tensor with a 
certain curvature tensor. In the case of Einstein-Hilbert gravity, this 
curvature tensor is the intrinsic Einstein-tensor of the metric while in our 
ansatz \eqref{eq:leftoverIsrael} this is an extrinsic curvature tensor. As 
pointed out in \cite{Curiel:2014zba} for example, certain energy conditions can 
be motivated from the matter side as conditions that realistic matter fields 
should satisfy, but others may be motivated from the curvature side. In our 
discussion of the SEC, we take the second approach. Especially in section 
\ref{sec::Kondo} we will find it phenomenologically very important to work with 
a matter content on the brane that does violate what we call SEC.}
\begin{align}
(S_{ij}-S\gamma_{ij})m^{i}m^{j}\geq0\ \  \forall\  m^i m_i<0.
\end{align}
The reason for this choice is that with $\gamma_i ^i=2$, the 
equation \eqref{eq:leftoverIsrael} can be rewritten as
\begin{align}
K^+_{ij}=-\frac{\kappa}{2} (S_{ij}-S\gamma_{ij}),
\end{align}
which will be useful in section \ref{sec::Wall}. In particular, using the 
decomposition \eqref{decompofS} again, the SEC takes the form (for optimal 
choice of $\alpha,\beta$):
\begin{align}
S_L\geq0,\ \ S_R\geq0,\ \ &S\geq -2 \sqrt{S_L S_R}.
\end{align}

The NEC is the most fundamental energy condition discussed here, as 
it is implied by both WEC and SEC. Although we will study WEC violating setups 
in section \ref{sec::toymodel} and SEC violating setups in sections 
\ref{sec::toymodel} and \ref{sec::Kondo}, we will not consider any NEC 
violations in this paper. Such NEC violations may be possible for exotic forms 
of classical matter \cite{Visser:1999de,Curiel:2014zba}, but  NEC is 
expected to hold in string theory, see \cite{Parikh:2014mja}. It is also 
possible to define other energy conditions (see 
e.g.~\cite{Curiel:2014zba,Martin-Moruno:2013wfa}), but these will not play a 
prominent role in this work. 

\subsection{Static case}
\label{sec::staticcase}

In this section, we consider the static case, in which the bulk-spacetime is 
assumed to be static with Killing time coordinate $t$ and where the embedding 
of the brane as well as the matter fields living on the brane are assumed to be 
independent of $t$.

From previous results \eqref{eq:exteriorcurvature} 
and \eqref{eq:pullbackMetric}, we get
\begin{equation}
 \begin{aligned}
 K^\pm_{tz}=K^\pm_{zt}&=0
 %\nonumber
%
 \end{aligned}
\end{equation}
and
\begin{equation}
 \gamma=
 \begin{pmatrix}
  g_{tt} & 0 \\
  0 	 & g_{zz}+\left(\partial_z x_{\pm}\right)^2 g_{xx}
 \end{pmatrix}.
\end{equation}

Hence, for the static symmetric case in the BTZ background we are going to 
investigate, the equation \eqref{eq:leftoverIsrael} requires $S_{ij}$ to be 
diagonal. Let us now write the induced metric as 
\begin{equation}
\gamma_{ij}= 
\begin{pmatrix}
  -a(z) & 0 \\
  0 	 & b(z)
 \end{pmatrix},
 \label{abmetric}
\end{equation}
with $a(z),b(z)>0$. The two null vectors $l^i$ and $r^i$ can easily be found, 
and the decomposition of $S_{ij}$ now reads
\begin{align}
 S_{ij}&=\frac{S}{2}\gamma_{ij}+S_{L}l_i l_j +S_{R} r_i r_j
 \\
 &=\frac{S}{2}
 \begin{pmatrix}
  -a & 0 \\
  0 	 & b
 \end{pmatrix}
 +\frac{S_L}{2}
 \begin{pmatrix}
  a & -\sqrt{ab} \\
  -\sqrt{ab} 	 & b
 \end{pmatrix}
  +\frac{S_R}{2}
 \begin{pmatrix}
  a & \sqrt{ab} \\
  \sqrt{ab} 	 & b
 \end{pmatrix}.
\end{align}
As staticity demands $S_{ij}$ to be diagonal in the coordinate system we chose 
in \eqref{abmetric}, this implies $ S_L=S_R\equiv S_{L/R}$ and hence 
\begin{align}
 S_{ij}=\frac{S}{2}
 \begin{pmatrix}
  -a & 0 \\
  0 	 & b
 \end{pmatrix}
 +S_{L/R}
 \begin{pmatrix}
  a & 0 \\
  0 	 & b
 \end{pmatrix}.
\end{align}
The physical meaning of this is simply that the matter content of the brane has 
to be at rest with respect to the Killing time direction $\partial_t$.

Let us now collect some formulae that will be helpful later on. The traceless 
symmetric tensor 
\begin{align}
 \tilde{\gamma}_{ij}\equiv \begin{pmatrix}
  a & 0 \\
  0 	 & b
 \end{pmatrix}
,\ \ 
 \tilde{\gamma}^{ij}= \begin{pmatrix}
  \frac{1}{a} & 0 \\
  0 	 & \frac{1}{b}
 \end{pmatrix},\ \ 
 \tilde{\gamma}_{ij} \tilde{\gamma}^{ij}=2
 \label{gammatilde}
  \end{align}
can be written as $\tilde{\gamma}_{ij}=\gamma_{ij}+2u_i u_j$ with the 
normalised timelike vector
$
 u_i=\begin{pmatrix}
  \sqrt{a}  \\
  0 
 \end{pmatrix}
$. 
It will also be of use later on to note that
\begin{align}
\nabla_i u^i&=0,
\label{nablauvanishes}
\\
 u^i\nabla_i u^j
&=\begin{pmatrix}
  0 \\
  \frac{1}{2}\gamma^{tt}\gamma^{zz}\partial_z \gamma_{tt}  
 \end{pmatrix}.
 \label{unablauvanishes}
\end{align}

The same decomposition as for $S_{ij}$ is also possible for the right 
side of the equation
\begin{equation}
 \mathcal{K}_{ij}\equiv-\left(K^+_{ij} - \gamma_{ij} 
K^+\right) = 
\frac{\kappa}{2}\,S_{ij},
\label{IsraelEOM}
\end{equation}
i.e. 
\begin{equation}
 \mathcal{K}_{ij}=\frac{\mathcal{K}}{2}
 \gamma_{ij}
+\mathcal{K}_{L/R}
\tilde{\gamma}_{ij},
\label{Kdecomp}
 \end{equation}
such that the equations of motion reduce to
\begin{align}
 \mathcal{K}=\frac{\kappa}{2}S\ \text{    and    
}\ \mathcal{K}_{L/R}=\frac{\kappa}{2}S_{L/R}.
  \label{EOMagain}
\end{align}
This form for the equations \eqref{eq:leftoverIsrael} is very interesting, as 
$S$ and $S_{L/R}$ are directly constrained by energy conditions. We summarise 
our results for \AdS and BTZ metrics of the form \eqref{BTZmetric} in table 
\ref{tab::table1}.

\begin{table}[htb]
\begin{center}
\begin{tabular}{r|l}
\hline
\multicolumn{2}{|c|}{BTZ $h(z)=1-z^2/z_H^2$, \qquad $z_H>z$}\\
\hline\\[-.3cm]
  $\mK_{L/R}$ & $\frac{z \left(z_H^2-z^2\right) \left(z 
x'_{+}{}^3+z_H^2 x''_{+}\right)}{2 z_H L
\left(z_H^2+\left(z_H^2-z^2\right) x'_{+}{}^2\right){}^{3/2}}$   \\[.4cm]
  $\mK$ & $\frac{2 z_H^4 x'_{+}+\left(z^4-3 z^2 z_H^2+2 z_H^4\right) 
x'_{+}{}^3+z z_H^2 \left(z^2-z_H^2\right) x''_{+}}{z_H L 
\left(z_H^2+\left(z_H^2-z^2\right) x'_{+}{}^2\right){}^{3/2}}$  \\[.4cm]
NEC ($\mK_{L/R}\geq0$) & $z x'_{+}{}^3+z_H^2 x''_{+}\geq0$\\[.3cm]
WEC ($2\mK_{L/R}-\mK\geq0$) &$z z_H^2 
\left(z_H^2-z^2\right) x''_{+}-z_H^4 
x'_{+}-\left(z_H^2-z^2\right)^2 x'_{+}{}^3 \geq0$\\[.3cm]
SEC ($2\mK_{L/R}+\mK\geq0$) & $x_{+}' \geq0$\\
\multicolumn{2}{}{}\\

\hline
\multicolumn{2}{|c|}{AdS $h(z)=1$}\\
\hline\\[-.3cm]
  $\mK_{L/R}$ & $\frac{z x_{+}''}{2 L
\left(1+x_{+}'{}^2\right)^{3/2}}$   \\[.4cm]
  $\mK$ & $\frac{2 x_{+}'+2 x_{+}'{}^3-z 
x_{+}''}{L\left(1+x_{+}'{}^2\right)^{3/2}}$  \\[.4cm]
NEC ($\mK_{L/R}\geq0$) & $x_{+}''\geq0$\\[.3cm]
WEC ($2\mK_{L/R}-\mK\geq0$) &$z 
x_{+}''-x_{+}'-x_{+}'{}^3 \geq0$\\[.3cm]
SEC ($2\mK_{L/R}+\mK\geq0$) & $x_{+}' \geq0$\\

\end{tabular}
\caption{Table summarising our findings of the impact of the different energy 
conditions NEC, WEC and SEC for static embeddings in BTZ and AdS backgrounds 
\eqref{BTZmetric}. Note that the NEC is part of the WEC and SEC, i.e.~WEC 
means that the NEC is satisfied and \textit{additionally} 
$2\mK_{L/R}-\mK\geq0$. In the first two lines each we give the extrinsic 
curvature scalars defined in \eqref{Kdecomp}. The last line in each case implies 
that the curve $x_+$ would bend to the right in our figure \ref{fig::geometry} 
iff the SEC where satisfied. Indeed, the general example shown in figure 
\ref{fig::geometry} contains violations of any energy condition for some value 
of $z$.
}
\label{tab::table1}
\end{center}
\end{table}

\subsection{Energy-momentum conservation}

Using the decomposition into scalars and the choice of coordinate system 
described in section \ref{sec::staticcase}, we describe in this subsection 
how to phrase the conservation of the hypersurface energy-momentum tensor 
$S_{ij}$ in a very simple and useful way. Remember that we are working in the 
static setting discussed in section \ref{sec::staticcase}. Conservation of the 
energy-momentum tensor $S_{ij}=S/2\gamma_{ij}+S_{L/R}\tilde{\gamma}_{ij}$ now 
demands, using $\nabla_i \gamma^{ij}=0$ and 
$\tilde{\gamma}_{ij}=\gamma_{ij}+2u_i u_j$,\footnote{This is implied by the 
Israel junction conditions \eqref{IsraelEOM} and the fact that $\nabla_i 
\mK^{ij}=0$ is a geometrical identity, at least for an embedding in an AdS or 
BTZ background. See \cite{Kwon:2012tp} for a related discussion.}
\begin{align}
 0=\nabla_i S^{ij}=\frac{1}{2}\partial_i S \gamma^{ij}+\partial_i S_{L/R} 
\tilde{\gamma}^{ij}+S_{L/R}\left(2 u^j\nabla_i u^i+2u^i\nabla_i u^j\right).
\end{align}
In this expression, we have $u^j\nabla_i u^i=0$. As the tensors 
$\gamma_{ij}$ and $\tilde{\gamma}_{ij}$ 
are diagonal, we can now easily investigate the two possible choices of the 
index $j$ above.

\paragraph{$j=t:$} As we assume staticity, i.e.~$\partial_t S=\partial_t 
S_{L/R}=0$, we end up with
$%\begin{align}
  0=\nabla_i S^{it}=2S_{L/R}u^i\nabla_i u^t
$ %\end{align}
which holds by \eqref{unablauvanishes}.

\paragraph{$j=z:$} Now, using again \eqref{unablauvanishes} and 
$\gamma^{zz}=\tilde{\gamma}^{zz}$, we find 
\begin{align}
 0&=\nabla_i 
S^{iz}=\frac{1}{2}S'\gamma^{zz}+S'_{L/R}\gamma^{zz}-\frac{2}{z}S_{L/R}\gamma^
{zz}
\\
&\Rightarrow \left(S+2S_{L/R}\right)'=\frac{4}{z}S_{L/R}
\label{consistencyAdS}
\end{align}
for a static embedding in AdS space (\eqref{BTZmetric} with $h(z)=1$) and 
\begin{align}
\left(S+2S_{L/R}\right)'=\frac{4}{z-\frac{z^3}{z_H^2}}S_{L/R}
\label{consistency}
\end{align}
in the BTZ case. Although the equation \eqref{consistency} is just the 
conservation of the energy-momentum tensor in our coordinate system, its simple 
form will be very useful in section \ref{sec::PF}. Also this equation has 
already a profound implication on the energy conditions: Note that (outside of 
the horizon) the right hand side of \eqref{consistency} is $\geq0$ by NEC, 
hence the quantity $S+2S_{L/R}$ inside the brackets can only grow with $z$. 
Comparing with table \ref{tab::table1}, this already tells us that assuming NEC, 
if the SEC is satisfied near the boundary, it cannot become violated deeper 
inside the bulk, as the SEC requires precisely $S+2S_{L/R}\geq0$. 

%%%%%%%%%%%%%%%%%%%%%%%%%%%%%%%%%%%%%%%%%%%%%%%%%%%%%%%%%%%%%%%%%%%%%%%%%%%%%%%%
%%%%%%%%%%%%%%%%%%%%%%%%%%%%%%%%%%%%%%%%%%%%%%%%%%%%%%%%%%%%%%%%%%%%%%%%%%%%%%%%

\section{Implications of energy conditions in two dimensions}
\label{sec::Wall}

\begin{figure}[ht!]
\centering
 \def\svgwidth{0.95\columnwidth}
\executeiffilenewer{drawing1.svg}{drawing1.pdf}%
{inkscape -z -D --file=drawing1.svg %
--export-pdf=drawing1.pdf --export-latex}%
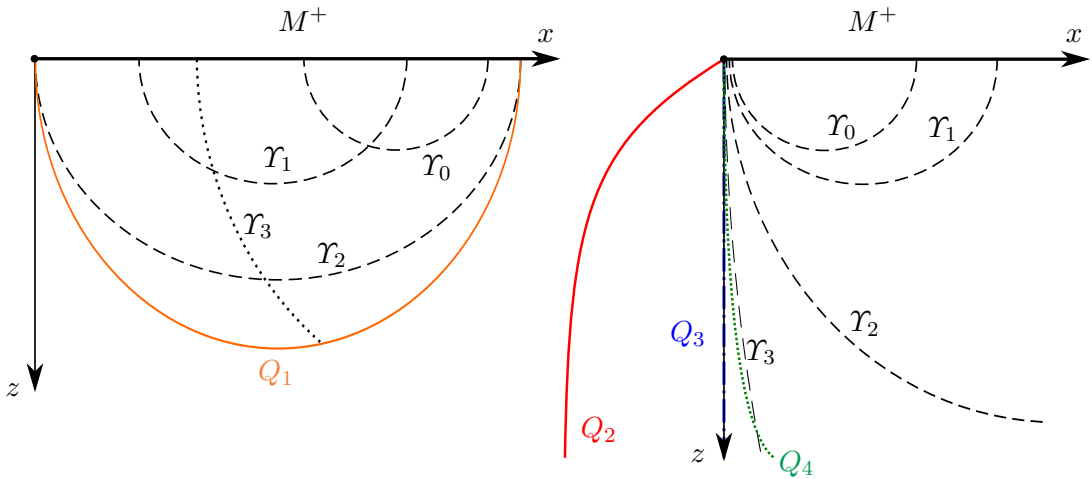%

\caption{
\textit{Left:} The case where the brane $Q_1$ bends back to the boundary. 
$\varUpsilon_0$, $\varUpsilon_1$, $\varUpsilon_2$ are $Q$-deformable spacelike 
extremal curves which are half-circles in a \Poincare background. The brane is 
crossed by $\varUpsilon_3$, but as $Q_1$ is anchored at the boundary 
\textit{twice}, $M^+$ is a finite interval, and as the point where 
$\varUpsilon_3$ returns to the boundary is not inside of $M^+$, it does not 
violate the barrier theorem. The matter fields sustaining $Q_1$ hence may 
satisfy WEC and SEC everywhere, and $Q_1$ is an extremal surface barrier.
\textit{Right:} Possible cases where branes $Q_i$ are anchored 
\textit{once} to the boundary $M^+$, which hence extends infinitely in one 
direction. Several $Q$-deformable extremal spacelike curves $\varUpsilon_a$ are 
depicted. $Q_3$ is a trivially embedded brane with $K^+_{ij}=0$ (i.e.~sustained 
by $S_{ij}=0$) and is an extremal surface barrier in the sense of the theorem 
presented above. $Q_2$ violates SEC as it has $x'_+(z)<0$ (see table 
\ref{tab::table1}), but as it falls behind $Q_3$ everywhere it is also an 
extremal surface barrier. $Q_4$ is crossed by $\varUpsilon_3$, and must 
according to the barrier theorem hence violate the WEC somewhere.}
\label{fig::drawing1}
\end{figure}

We now investigate the relation between the brane and geodesic curves, and show 
that when the WEC and SEC are satisfied, the brane bends back to the boundary. 
As we will see for a toy model in section \ref{sec::toymodel}, it is of 
importance whether geodesics that both start and end at the boundary to the same 
side of the brane will reach the brane, see the left drawing in figure 
\ref{fig::drawing1}. In this context, it will be enlightening to consider the 
theorems presented by Engelhardt and Wall in \cite{Engelhardt:2013tra}. These 
authors proved that under certain conditions a codimension one hypersurface 
is an \textit{extremal surface barrier}, i.e.~a surface such that spacelike 
extremal surfaces anchored to one side of the barrier cannot cross it. For 
example, in section \ref{sec::toymodel} where we will investigate the case of a 
brane with constant tension $\lambda$, the brane is such a barrier for 
$\lambda>0$, but not for $\lambda<0$.

Let us now first state the theorem and then relate its assumptions to the 
energy conditions on the hypersurface. We will phrase the theorem in 
terms of our nomenclature introduced in figure \ref{fig::geometry}.

\textbf{\textit{Barrier Theorem}} (Engelhardt, Wall 
\cite{Engelhardt:2013tra})\footnote{There are some 
subtleties related to whether an extremal surface may touch the extremal 
surface barrier or not, which are not of much importance to us in this work. 
Hence the theorem presented here is a combined result of theorems 2.1, 2.2 and 
corollary 2.4 in \cite{Engelhardt:2013tra}.} \\
Let $Q$ be a hypersurface splitting the spacetime $N$ in two parts ($N_+$ and 
$N_-$) such that $K^{+}_{ij}v^{i}v^{j}\leq0$ for any vector field $v^{i}$ on 
$Q$. Then any $Q$-deformable\footnote{This means that we assume there to be a 
family $\{\varUpsilon_a\}$ of extremal spacelike surfaces such that all of 
those are anchored on $M_+$, and can be continuously deformed from some 
$\varUpsilon_0\in\{\varUpsilon_a\}$ which is only located in $N_+$ and does 
not touch $Q$. We refer to any member $\varUpsilon$ of this family as
\textit{$Q$-deformable}.} 
spacelike extremal surface $\varUpsilon$ which is anchored in $M_+$ remains 
in $N_+$.
\\

This means that if we have two points $A$ and $B$ on the boundary (to the same 
side of the brane $Q$), the extremal curve $\varUpsilon_{[AB]}$ connecting the 
two points in the bulk cannot cross $Q$ if the extrinsic curvature on $Q$ 
satisfies the assumption made in the theorem.

This theorem is also of relevance for holographic computations of entanglement 
entropy. In the Ryu-Takayanagi proposal \cite{Ryu:2006bv,Ryu:2006ef}, the 
entanglement entropy of some spatial area on the field theory side is 
proportional to the area of extremal surfaces in the bulk. Hence, in view of the 
holographic calculation of entanglement entropy in the present context, it is 
interesting to know whether the hypersurface $Q$ is a barrier surface as defined 
above or not. 

As the barrier theorem relies on an assumption concerning the extrinsic 
curvature tensor $K_{ij}$, we use the equations \eqref{eq:leftoverIsrael} to 
relate this assumption to properties of the energy-momentum tensor $S_{ij}$. The 
issue is that the condition utilised above is supposed to hold for \textit{any} 
vector field $v^{i}$, which includes spacelike vectors. Although energy 
conditions usually only restrict the contraction of $S_{ij}$ with causal vectors 
(see section \ref{sec::ECs2}), we now show that the assumption in the above 
theorem is indeed implied by particular energy conditions. This provides us with 
another way to determine qualitatively how the brane may bend (see figure 
\ref{fig::drawing1}), which is based on whether the energy conditions are 
satisfied or violated by matter fields localised on $Q$.

As a next step, we relate the assumptions of the barrier theorem to typical
energy conditions in 1+1 dimensions by applying the Israel junction conditions.
From \eqref{eq:leftoverIsrael}\footnote{As we are only 
interested in energy conditions, we set the positive prefactor 
$\frac{\kappa}{2}$ to one from now on.}
it follows that $K=S$ (dropping the superscript $+$ from now on) and hence
\begin{align}
 -K_{ij}=+S_{ij}-S\gamma_{ij}.
 \label{Wall}
\end{align}
Demanding $K_{ij}v^i v^j\leq0$ hence corresponds to 
$S_{ij}v^i v^j-S v_i v^i\geq0$,
which with the usual decomposition of $S_{ij}$ reads
\begin{align}
 -\frac{S}{2}v_i v^i + S_L (l_i v^i)^2 + S_R (r_i v^i)^2\geq0.
\end{align}
The last two terms are nonnegative by positivity of the squares and the NEC. 
There are now three cases: $S>0$, $S=0$ and $S<0$.

In the case $S=0$, the above inequality is satisfied for any $v^i$ due to the 
NEC. $S=S_L=S_R=0$ would be the trivial 
case. 

In the case $S< 0$, the term $-\frac{S}{2}v_i v^i$ can only be negative 
(i.e.~problematic) for timelike vectors $v_i$, for which the above equation 
corresponds to the SEC.

In the case $S>0$, the term $-\frac{S}{2}v_i v^i$ can only be negative 
(i.e.~problematic) for spacelike vectors $v_i$. Assuming a decomposition 
$v^i=-\alpha l^i+\beta r^i$ (with $\alpha\cdot\beta>0$) we find 
\begin{align}
 -S\alpha\beta+S_L \beta^2+S_R\alpha^2\geq0\ \ \text{for any 
$\alpha\cdot\beta>0$}
\end{align}
which corresponds to the WEC.

Hence we see very generally that the condition of the barrier theorem 
corresponds to SEC and WEC being satisfied simultaneously,\footnote{
Remember that NEC is implied by WEC, see 
section \ref{sec::ECs2}.} which is a rather strict condition on the 
energy-momentum tensor. In this case, the brane $Q$ is an extremal surface 
barrier in the sense of \cite{Engelhardt:2013tra}, but there may be extremal 
surface barriers for which SEC or WEC are violated. Especially in sections 
\ref{sec::toymodel} and \ref{sec::Kondo} we will encounter examples 
where SEC is violated, but where the brane is located behind an extremal surface 
barrier and is hence an extremal surface barrier itself, see also figure 
\ref{fig::drawing1}. 

The barrier theorem described above allows for a nice and simple corollary that 
tells us how certain branes bend in the bulk spacetime depending on the energy 
conditions that the matter content of the brane satisfies or violates. Suppose 
that we have a brane where near the boundary the SEC is satisfied, which by 
table \ref{tab::table1} means that $x'_+(z)\geq0$ for small enough $z$. By the 
conservation of energy-momentum \eqref{consistency} we know that then SEC will 
be satisfied everywhere when NEC holds. Qualitatively, the brane hence looks 
like either $Q_1$ or $Q_4$ in figure \ref{fig::drawing1}. The case $Q_1$ is 
possible for the situation where WEC and SEC are satisfied everywhere on the 
brane. We will obtain precisely this kind of behaviour is section \ref{sec::PF}. 
For the case $Q_4$ we see that an extremal curve $\varUpsilon_3$ crossing the 
brane can indeed easily be constructed: Take a curve $\varUpsilon_0$ that 
connects boundary points $x_2>x_1>0$ in $M^+$ defined by $x\geq0$. The larger 
$x_2-x_1$, the deeper the curve enters into the bulk. When now sending 
$x_1\rightarrow 0$ and $x_2\rightarrow\infty$ (which is not possible if the 
brane reaches back to the boundary, such as $Q_1$), we will sooner or 
later encounter an intersection between the brane and the geodesic. Hence, by 
the barrier theorem a situation as sketched with $Q_4$ in figure 
\ref{fig::drawing1} can only appear if somewhere on $Q_4$, WEC and/or SEC are 
violated. Phrased differently, this means that if WEC and SEC are satisfied 
everywhere by the matter content on the brane $Q$, it has to bend over and 
return to the boundary\footnote{Exluding, of course, the case of a trivial 
embedding with $K^+_{ij}=0\Rightarrow S_{ij}=0$.}. 

The results summarised in table \ref{tab::table1} already point at this for the 
case of a BTZ black hole as background, however the barrier theorem gives a 
more general argument. Assuming that the brane enters the event horizon 
with $x_+'(z_H),x_+''(z_H)$ finite, we find that WEC implies $x_+'(z)\leq0$, 
which means that (again except for the trivial case $K^+_{ij}=0$) either SEC or 
WEC have to be violated. If they are not violated, the brane has to turn around 
before reaching the event horizon and return to the boundary, see section 
\ref{sec::PFinBTZ} for examples. In section \ref{sec::Kondo}, we will present a 
system where for model building purposes it is very important that SEC is 
violated everywhere on $Q$, giving a geometry similar to $Q_2$ in figure 
\ref{fig::drawing1}. 

It is a natural question to ask whether this relation between WEC and 
SEC and the assumption in the barrier theorem generalises to higher 
dimensions. We will briefly comment on this in appendix \ref{sec::Wall2}. We 
find that the results presented here do not easily generalise to higher 
dimensions, meaning that the case of a two-dimensional hypersurface $Q$ is very 
special, in that here the satisfaction or violation of the WEC and SEC will have 
profound and immediate consequences for the geometry of the hypersurface. 
Indeed, in later sections \ref{sec::PF} and \ref{sec::Kondo} we will see 
concrete examples where the geometry of the hypersurface $Q$ is determined by 
the energy conditions that the matter fields living on this hypersurface do or 
do not satisfy, just as explained in this section for the general case.

%%%%%%%%%%%%%%%%%%%%%%%%%%%%%%%%%%%%%%%%%%%%%%%%%%%%%%%%%%%%%%%%%%%%%%%%%%%%%%%%
%%%%%%%%%%%%%%%%%%%%%%%%%%%%%%%%%%%%%%%%%%%%%%%%%%%%%%%%%%%%%%%%%%%%%%%%%%%%%%%%
  
\section{Special case: Constant brane tension}
\label{sec::toymodel}

Before studying situations with non-trivial matter fields on the hypersurface 
$Q$, in this section we revisit a simple model that has already been 
investigated in \cite{Azeyanagi:2007qj}, namely the model where we have a 
brane with constant tension embedded into global $\text{AdS}_3$. This will 
serve as a first example for applying the results of sections \ref{sec::ECs} 
and \ref{sec::Wall}. The case of a brane with a constant tension embedded in 
\Poincare AdS is very simple and will be discussed as a further application in 
section \ref{sec::PF+C}.

\subsection{Toy model: Global AdS}

In \cite{Azeyanagi:2007qj}, a toy model was studied which involves a 
brane $Q$ with action\footnote{See also section 2 of \cite{Karch:2000gx} for 
earlier thoughts in this direction.} 
\begin{align}
 \mS=-\lambda\int d^2x\sqrt{-\gamma}
 \label{consttension}
\end{align}
embedded in global $\text{AdS}_3$. In \eqref{consttension}, $\gamma$ is the 
determinant of the induced metric on the 1+1-dimensional brane. $\lambda$ 
will be referred to as \textit{brane tension}, although out of mathematical 
curiosity we will also consider negative values $\lambda<0$. $\lambda$ may 
be considered as a cosmological constant living on the brane. 

The authors of \cite{Azeyanagi:2007qj} worked with a very useful coordinate 
system with coordinates $t,r,y$ and %metric
line element
\begin{align}
ds^2=-\cosh(r)^2\cosh(y)^2dt^2+\cosh(y)^2dr^2+dy^2,
\label{AdSmetric}
\end{align}
where the AdS scale is set to $L=1$. This is related to the usual coordinates 
$t,\rho,\phi$ with %metric
line element
\begin{align}
ds^2=-\cosh(\rho)^2 dt^2+d\rho^2+\sinh(\rho)^2 d\phi^2
\end{align}
via the relations $\cosh(y)\cosh(r)=\cosh(\rho)$, 
$\sinh(y)=\sinh(\rho)\sin(\phi)$.

\subsection{Brane embedding}
\label{sec::toymodelbraneembedding}

\begin{figure}[htb]
\begin{center}
 \includegraphics[width=0.6\textwidth]{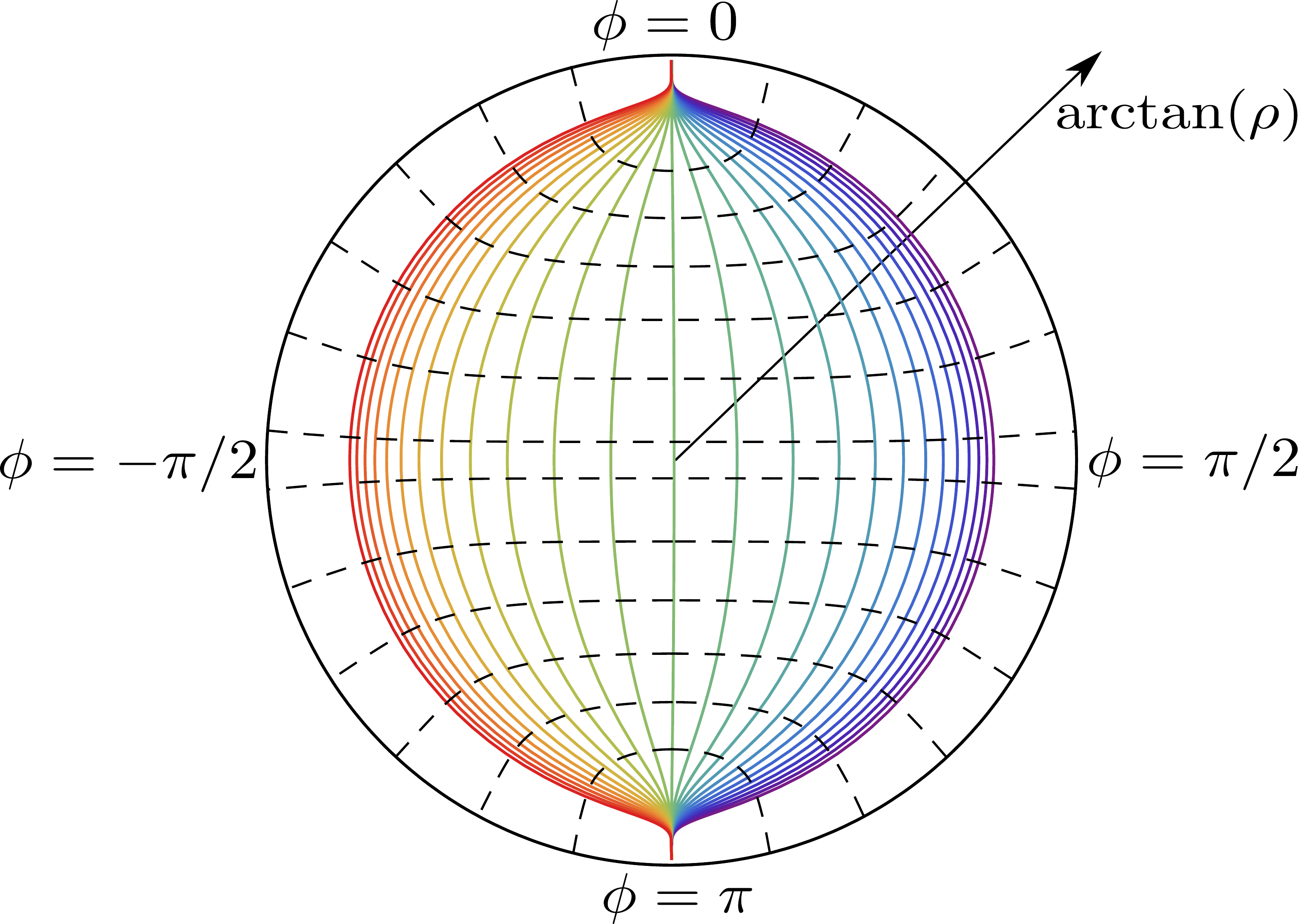}
\caption{Embedding of the branes given by $y=-y_*$ in AdS 
space with coordinates $t,\rho,\phi$. The spacetime to the right of each of 
these curves corresponds to the region $N_+$ in figure \ref{fig::geometry}, 
while the part of the spacetime to their left is excised. We set $t=const.$ to 
consider a spacelike slice with coordinates $\rho,\phi$, and compactify by 
plotting $\arctan(\rho)$ as radial coordinate, such that the thick black circle 
represents the AdS boundary. $\phi$ is the angular coordinate. The values 
of $\lambda$ for the branes shown are specified by 
$y_*=\arctanh(\frac{\kappa\lambda}{2})=\{-3,-2.75,...,3\}$. Lines involving 
red colour (bending to the left) stand for branes with $\lambda>0$, while  lines
involving blue colour (bending to the right) stand for branes with $\lambda<0$. 
The straight vertical line is the brane with zero tension, $\lambda=0$. The 
black dashed lines are geodesics perpendicular to the branes, see section 
\ref{sec::normalgeodesics}.
}
\label{fig::branes}
\end{center}
\end{figure}

\begin{figure}[htb]
\begin{center}
 \includegraphics[width=0.6\textwidth]{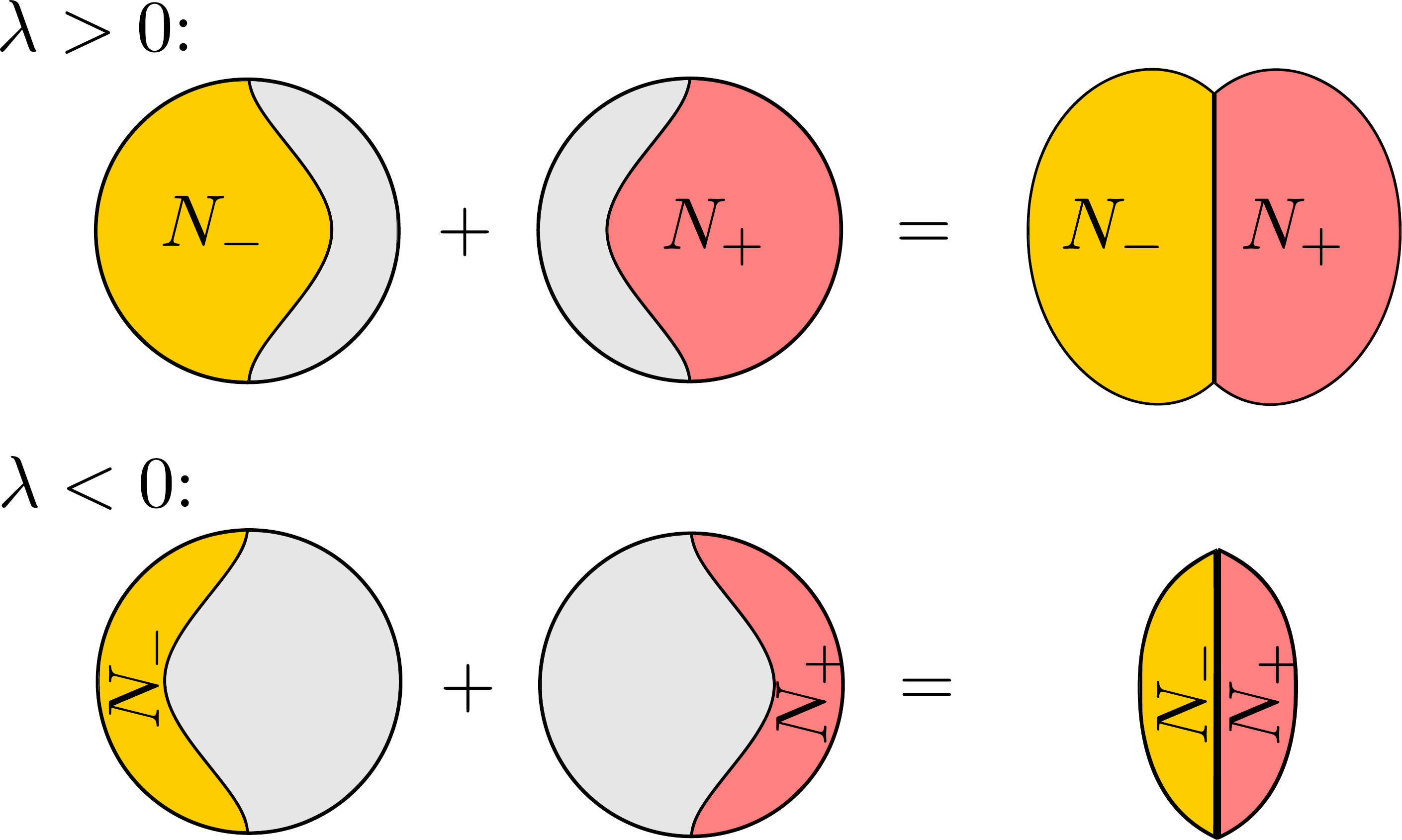}
\caption{Nontrivial matching of two AdS spacetimes along a 
constant tension brane for the cases $\lambda>0$ and $\lambda<0$. The two 
figures to the left of the equality depict the embedding of the brane with 
respect to $N_-$ and $N_+$, respectively. The grey shaded area is then 
excised, and the two resulting spacetimes are glued together along the brane, as 
shown to the right of the equality. By assumption of symmetry, $N_-$ is 
always the mirror image of $N_+$. As described in the text, the resulting 
spacetime will have increased volume for $\lambda>0$ and smaller volume than 
\AdS for $\lambda<0$.
}
\label{fig::AdSmatching}
\end{center}
\end{figure}

In \cite{Azeyanagi:2007qj} it was found that the embedding of the 
brane is given by two AdS spaces sewed together along lines of constant 
$y$, see figures \ref{fig::branes} and \ref{fig::AdSmatching}. 
We will now verify these results using the Israel junction conditions governing 
the behaviour of the brane. To do so, similarly to the setup presented in 
figure \ref{fig::geometry}, we will assume that in $N_+$ the embedding of the 
brane is given by a function $y_+(r)$, which we will then show to be constant.

From \eqref{consttension} we immediately obtain
\begin{align}
 S_{ij}=-\frac{2}{\sqrt{-\gamma}}\frac{\delta\left(-\lambda\sqrt{
-\gamma}\right)} { \delta\gamma^ { ij } } =-\lambda \gamma_{ij},
\end{align}
with induced metric $\gamma_{ij}$. Note that for $\lambda\geq0$ this satisfies
the WEC, see section \ref{sec::ECs}. The equations 
\eqref{eq:leftoverIsrael} then read
\begin{align}
 -K_{ij}+K\gamma_{ij}=-\frac{\kappa \lambda}{2}\gamma_{ij}.
 \label{simplifiedequations}
\end{align}
Assuming the brane embedding $y_+(r)=-y_*=const.$ we find
\begin{align}
n^{\mu}&=\delta^\mu_{3},\\
K_{ij}&=
\left(
\begin{array}{cc}
 \cosh(y_*)\sinh(y_*) \cosh(r)^2 & 0 \\
 0 & -\cosh(y_*)\sinh(y_*) \\
\end{array}
\right),
\\
\gamma_{ij}&=
\left(
\begin{array}{cc}
 -\cosh(y_*)^2 \cosh(r)^2 & 0 \\
 0 & \cosh(y_*)^2 \\
\end{array}
\right).
\end{align}
From this and \eqref{simplifiedequations} it follows that 
\begin{equation}
 \tanh(y_*)=\frac{\kappa\lambda}{2}=4\pi G_N \lambda
 \label{eq::TakayanagiGeodesicLengthAndBraneTension}
\end{equation}
precisely as in \cite{Azeyanagi:2007qj}, i.e. the solutions to 
\eqref{simplifiedequations} are indeed given by a brane embedding of the form 
$y_+(r)=-y_*$. Obviously, there is an upper bound on the absolute value 
of the brane tension $\lambda$.

Let us now discuss the geometries obtained by this construction. For 
$\lambda=0=y_*$, i.e.~the case without tension, we just cut the AdS space in 
half and match the two sides (i.e.~$N_-$ and $N_+$, see figure 
\ref{fig::geometry}) trivially. For general values of $\lambda$, the profile of 
the brane $y=-y_*$ with respect to $N_+$ is shown in figure \ref{fig::branes}. 
$N_+$ in this figure is always to the right side of the brane. As we can see, 
for $\lambda>0$ the branes extend to the left, such that $N_+$ is larger than 
half of the AdS space, while for $\lambda<0$ (and hence WEC violation) the 
branes extend to the right, hence making $N_+$ smaller than half of the AdS 
space, see also figure \ref{fig::AdSmatching}.

\subsection{Normal geodesics}
\label{sec::normalgeodesics}

As seen in the previous section, the branes parameterised by their tension 
$\lambda$ are described by an embedding of the form $y=-y_{*}(\lambda)$. 
Obviously, the curves $x^{\mu}(y)=\left(t_0,r_0,y \right)$ are normal to all of 
these branes as $\dot{x}^{\mu}=n^{\mu}$ with the normal vector $n$ and have the 
coordinate $y$ as affine parameter. As it turns out, these curves actually 
describe geodesics, as with this ansatz the geodesic equations 
$\ddot{x}^{\mu}+\Gamma^{\mu}_{\alpha\beta}\dot{x}^{\alpha}\dot{x}^{\beta}=0$ 
simplify to $\Gamma^{\mu}_{yy}=0$. From the metric \eqref{AdSmetric} it is 
indeed easy to see that this is satisfied. Indeed, these geodesics normal to 
the branes are exactly those which are symmetric to reflections about the brane 
with vanishing tension, i.e.~with $\lambda=0\Rightarrow y_{*}=0$, see figure 
\ref{fig::branes}. Hence we know that due to the backreaction of the brane, all 
these geodesics are extended by an amount of $2y_*$ compared to the pure \AdS 
case $\lambda=0$. For boundary regions that extend symmetrically to the left 
and right from the point where the brane meets the boundary, this means that 
the entanglement entropy increases by an amount of exactly 
$y_*/2G_N$. This coincides with the result of \cite{Azeyanagi:2007qj}. 
In the next section, we will discuss entanglement entropy for intervals that do 
not include the point where the brane is anchored at the boundary.

\subsection{Entanglement entropy of general intervals}
\label{sec::furtherconsid}

The major motivation to investigate the backreaction on the geometry is that it
is necessary for calculating the entanglement entropy using the Ryu-Takayanagi 
proposal \cite{Ryu:2006bv,Ryu:2006ef}. This states that the entanglement entropy 
of some spacelike area in the field theory is proportional to the area of a 
spacelike minimal surface in the bulk with the same boundary.

In 2+1 bulk dimensions, the area will be given by a line segment $[AB]\subset M$
on the conformal boundary and the extremal surface by the geodesic connecting 
the boundaries of $[AB]$ which we denote $\varUpsilon_{[AB]}$. Furthermore 
$\partial [AB] = \partial \varUpsilon_{[AB]} = \{A,B\}$. The proposal is then 
given by

\begin{equation}
 S_{[AB]} = \frac{\text{min area}(\varUpsilon_{[AB]})}{4 G_N}.
 \label{RyuTakayanagi}
\end{equation}

In the probe limit, the geometry does not change according to the fields on the
defect and thus the entanglement entropy cannot be affected since localised
energy-momentum does not alter the variational problem for minimal surfaces.

In this section we point out one important difference between the cases of 
positive and negative brane tension $\lambda$ with respect to the 
holographic calculation of entanglement entropy. These cases can be physically
distinguished by WEC (satisfied for $\lambda\geq0$, violated for $\lambda<0$) 
and SEC (violated for $\lambda>0$, satisfied for $\lambda\leq0$).

\begin{figure}[htb]
\begin{center}
 \includegraphics[width=0.4\textwidth]{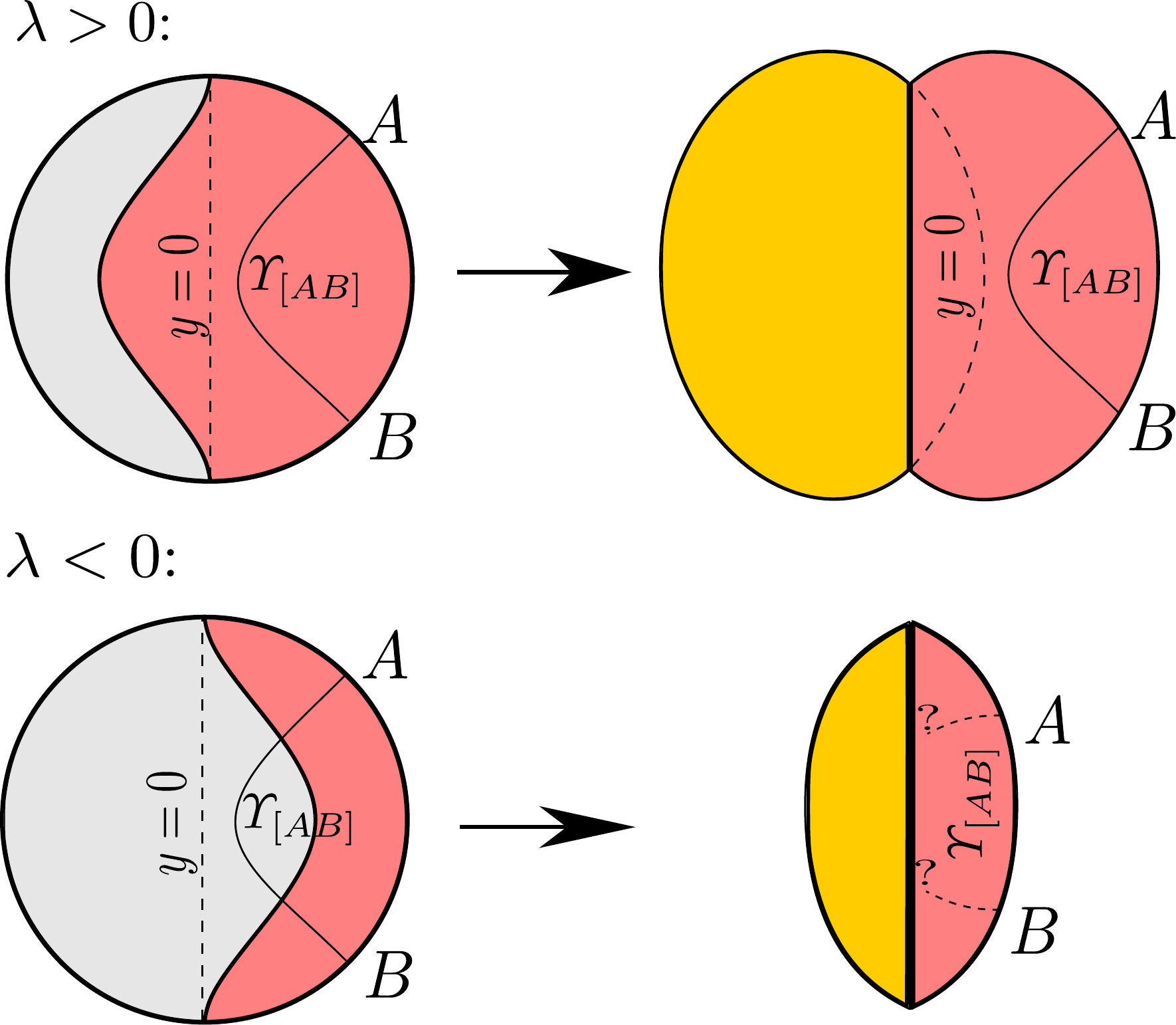}
\caption{Spacelike geodesics $\varUpsilon[AB]$ defining entanglement entropy 
for intervals $[AB]$ in the cases $\lambda>0$ and $\lambda<0$. $y=0$ would be 
the brane for the tensionless case $\lambda=0$. For $\lambda<0$, the geodesics 
$\varUpsilon[AB]$ may intersect the brane and may be reflected/refracted by it.
}
\label{fig::furtherconsid}
\end{center}
\end{figure}

First, let us mention that the surface defined by $y=0$ in figure 
\ref{fig::furtherconsid} is an \textit{extremal surface barrier} in the sense of 
\cite{Engelhardt:2013tra} and the barrier theorem given in
section \ref{sec::Wall}. This means that if we 
have two points $A$ and $B$ on the boundary (to the same side of the brane), the 
extremal curve $\varUpsilon_{[AB]}$ connecting the two points in the bulk cannot 
cross this surface, see figure \ref{fig::furtherconsid}. Note that as we work 
in $2+1$ bulk dimensions, and as we suppress the (static) time direction, 
the length of this curve $\varUpsilon_{[AB]}$ determines the entanglement 
entropy of the interval between $A$ and $B$.

For $\lambda>0$, the brane, given by $y=-y_*$,
stays behind the extremal surface barrier $y=0$ (see 
figure \ref{fig::furtherconsid}) and is hence itself an extremal surface 
barrier, just as $Q_2$ in figure \ref{fig::drawing1}. This means that the new 
spacetime we obtain has more volume than global 
$\text{AdS}_3$\footnote{Assuming a suitable regularisation.} and geodesics 
crossing the brane perpendicularly are longer than in the tensionless 
case, i.e.~the entanglement entropy given by these curves increases, see 
section \ref{sec::normalgeodesics}. This makes sense if we assume that the 
interface described by the brane introduces new degrees of freedom into the 
system. Geodesics which are not crossing the barrier, such as 
$\varUpsilon_{[AB]}$ in the figure \ref{fig::furtherconsid}, will be unaltered, 
which means that if we take any boundary interval $[AB]$ such that the brane 
does not reach the AdS boundary within this interval, its entanglement entropy 
will be precisely the same as in the pure AdS case.
 
For $\lambda<0$ in contrast, the brane at $y=-y_*$ crosses the extremal surface 
barrier $y=0$, see figure \ref{fig::furtherconsid}. This means that the new 
spacetime we obtain by excising the grey area has less volume than AdS, and 
geodesics crossing the brane perpendicularly are shorter than in the tensionless 
case, i.e.~also the entanglement entropy defined by these curves is smaller 
(see again section \ref{sec::normalgeodesics}). Even geodesics which are not 
crossing the barrier $y=0$, such as $\varUpsilon_{[AB]}$ in the figure 
\ref{fig::furtherconsid} may be cut off at the brane, hence the entanglement 
entropies of intervals like $[AB]$ may be described by altered curves. In 
principle, we expect that there should always be a minimal curve connecting any 
two points in the spacetime (see however \cite{Fischetti:2014uxa} for 
geometries where this is not the case), but for $\lambda<0$ it appears 
that there exist curves which must be refracted (or reflected) at the brane.

The refraction conditions at the brane follow from a local minimisation problem,
see \cite{JLM:146195}.
However, there is a subtlety:
Suppose we would like to connect two points $A$ and $B$ which lie in 
the interior 
(or on the conformal boundary) of $N_+$ for $\lambda<0$. Moreover, suppose that 
a segment of 
the original geodesic connecting the points is excised by our approach as 
indicated in figure \ref{fig::furtherconsid}.
In this case, the curve of minimal length connecting $A$ and $B$ clearly cannot 
lie only in 
the interior of $N_+$, since there the geometry is smooth and there is always a 
mean curvature flow which leads towards the original geodesic 
and hence the brane.
If the geodesic approaches the brane at a finite angle, then due to the 
refraction conditions (which are given by a generalised Snell's law, see
\cite{JLM:146195}) it would continue in $N_-$ with a finite angle, too.
In order to return to $N_+$ it must cross the brane at least once more.
However, the segment connecting both crossing points, regarded as elements of 
$N_-$, cannot be of minimal length for negative tension.
In the original spacetime, the geodesic connecting those points would lie 
entirely in the excised part.
The same argument holds for curves connecting the crossing points (now regarded 
as elements of $N_+$) and lying in the interior of $N_+$.
Hence the minimal curve connecting the ``crossing'' points must lie on the
brane. 
Thus we can reduce the problem to considering minimal curves lying in $N_+ 
\cup Q$ only.

\begin{figure}[htb]
\begin{center}
 \includegraphics[width=0.9\textwidth]{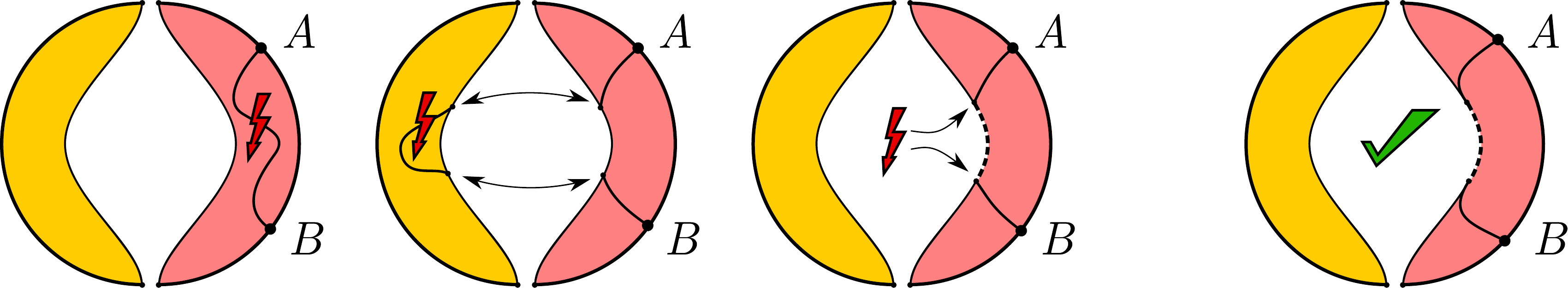}
\caption{The four possibilities of minimal curves connecting two points 
$A$ and
$B$ for $\lambda<0$ as considered in the main text. 
The first three possibilities are excluded
by general properties of geodesics in AdS if a part of the original geodesic is
excised by the approach. The only possibility left is the forth in which the 
dashed part of the curve is not geodesic w.r.t.~to the ambient geometry, but 
nevertheless minimal.}
\label{fig::furtherconsid2}
\end{center}
\end{figure}

Now that a finite segment of the minimal curve connecting $A$ and $B$ lies 
completely on the brane, the curve is not allowed to reach the brane
at a finite angle. 
Otherwise, it would have a kink at that point which can always be replaced by 
a smooth edge of minor length.
Thus the desired minimising curve connecting $A$ and $B$ consists of three 
segments:
The first segment is a geodesic in $N_+$ connecting $A$ with the entrance point 
on the brane at which it ends tangentially.
The second segment lies entirely on the brane, connects the 
entrance and exit points and is required to be a geodesic 
w.r.t.~the induced geometry on the brane.\footnote{As the segment lies 
entire on the brane, the notion of being a geodesic w.r.t.~the induced 
geometry on $Q$ is sensible. Regarded as a curve in $N_+$, this part of the 
curve cannot be geodesic w.r.t.~$N_+$ since the embedding of $Q$ is not totally 
geodesic for $\lambda \neq 0$. 
Nevertheless, this segment would be the minimal curve connecting the entrance 
and exit points on the brane because proper geodesics connecting those points 
and lying in $N_+$ or $N_-$ are excised by the approach and being a geodesic 
on $Q$, the curve is the minimal one allowed.} 
The last segment is again a geodesic in $N_+$ which leaves the brane 
tangentially and connects the exit point on the brane and $B$.
The discussion is illustrated in figure \ref{fig::furtherconsid2}.

To summarise, for $\lambda<0$ we obtain that the holographic entanglement 
entropy of intervals which do not include the defect itself becomes affected.
When tuning $\lambda$ from $0$ to $-\frac{1}{4\pi G_N}$, the first geodesics 
to intersect the brane are those reaching deep into the bulk and hence belonging 
to large intervals $[AB]$. The geodesics belonging to smaller intervals $[AB]$ 
will intersect the brane for smaller values of $\lambda$. 
Due to this behaviour, it would be very interesting to find systems in which 
$\lambda<0$ is satisfied and study the impact of what was found above, see also 
the discussion in section \ref{sec::conc}.

%%%%%%%%%%%%%%%%%%%%%%%%%%%%%%%%%%%%%%%%%%%%%%%%%%%%%%%%%%%%%%%%%%%%%%%%%%%%%%%%
%%%%%%%%%%%%%%%%%%%%%%%%%%%%%%%%%%%%%%%%%%%%%%%%%%%%%%%%%%%%%%%%%%%%%%%%%%%%%%%%

\section{Perfect fluid on the brane}
\label{sec::PF}

Let us now come to one of the main results in this paper. Using the 
decomposition of the equations \eqref{eq:leftoverIsrael} into scalar equations 
outlined in section \ref{sec::decomposition}, it is possible to find simple 
analytical expressions for the (static) embedding of the brane 
$Q$ in the case where the matter content on $Q$ is a perfect fluid.
We compute the brane embeddings explicitly for this case.

\subsection{Perfect fluid in \Poincare AdS}
\label{sec::PF1}

Let us consider the configuration where the brane matter is given by a perfect 
fluid. The brane is embedded in a \Poincare AdS geometry, hence $h(z)=1$ and 
$h'(z)=0$ in \eqref{BTZmetric}. To keep track of the AdS-scale $L$, we assume 
\begin{align}
 x_+(z)\rightarrow L\cdot x_+\left(\frac{z}{L}\right)
\end{align}
so that $x_+$ is a dimensionless function with a dimensionless argument, and 
find for the extrinsic curvature scalars defined in \eqref{IsraelEOM} and 
\eqref{EOMagain}
\begin{align}
 \mK_{L/R}&=\frac{z x_{+}''(z/L)}{2L^2(1+ x_{+}'(z/L)^2)^{3/2}},
\\
\mK&=\frac{2 L x_{+}'(z/L)^3+2L x_{+}'(z/L)-z 
x_{+}''(z/L)}{L^2(1+ x_{+}'(z/L)^2)^{3/2}}.
\end{align}

As we assume the matter content on the brane to be a perfect fluid, the 
energy-momentum tensor is given by\footnote{Recently, perfect fluids in an 
AdS/BCFT setting were studied in \cite{Magan:2014dwa}. Apart from 
working in one dimension less in this paper, another important difference is 
that we can assume the equation of state of the fluid as a matter of choice, 
while in \cite{Magan:2014dwa} the equation of state is a necessary consequence 
of the ansatz. Also, \cite{Magan:2014dwa} investigates a connection between 
AdS/BCFT and fluid/gravity correspondence, which is not our intent in this work.
}
\begin{align}
 S_{ij}=(\rho+p)u_i u_j +p \gamma_{ij}
\end{align}
with $u_i\sim(1,0)$ for staticity, which easily gives
\begin{align}
 S&=p-\rho,\ \ S_{L/R}=\frac{p+\rho}{2}.
 \label{PFscalars}
\end{align}
Let us now assume an equation of state\footnote{Also, solutions may exist for 
more complicated equations of state, and we found 
that for the form $p=a\cdot\rho^b$ at least part of the following computations 
can still be performed analytically.} 
\begin{align}
 p=a\cdot\rho
 \label{PFeos}
\end{align}
with constant $a$ so that 
\begin{align}
 S&=\rho(a-1),\ \ S_{L/R}=\rho\frac{1+a}{2}
 \label{PFscalarsa}
\end{align}
and the energy conditions, assuming $\rho>0$, read
\begin{align}
 \text{NEC:\  \ }a\geq-1, \ \ \text{WEC:\ \ }a\geq-1,\ \ \text{SEC:\ \ }a\geq0.
\end{align}

There are now two possible ways to proceed towards exact solutions of this 
system. The first one is to take the two scalar equations \eqref{EOMagain} and 
get rid of $\rho(z)$ by setting up the equation 
\begin{align}
 \frac{\mK}{\mK_{L/R}}=\frac{2(a-1)}{a+1}=const.
\end{align}
which is then a first order ODE for $x_{+}'(z/L)$ and can be easily 
solved by separation of variables. 

\begin{figure}[htb]
\centering
 \def\svgwidth{0.90\columnwidth}
\executeiffilenewer{HyperHyperAB.svg}{HyperHyperAB.pdf}%
{inkscape -z -D --file=HyperHyperAB.svg %
--export-pdf=HyperHyperAB.pdf --export-latex}%
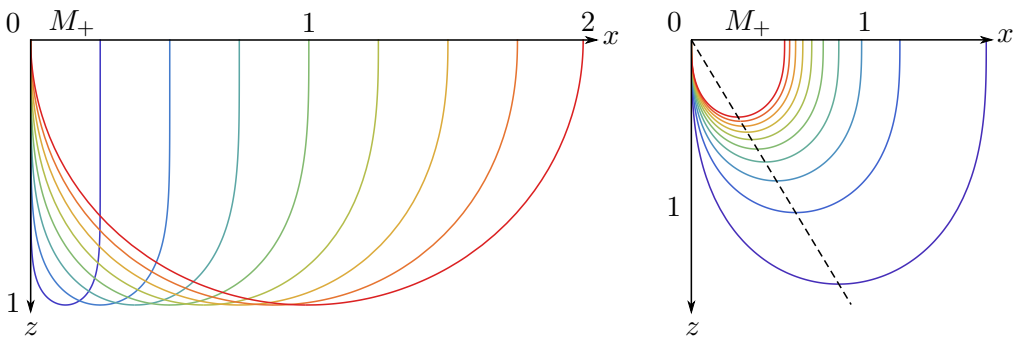%

\caption{\textit{Left:} Brane profiles \eqref{Hypergeo} (added with symmetric 
parts leading back to the boundary) for $L=1,c=1/a$ and, in order of increasing 
second intersection with the $x$-axis, 
$a\approx0.1,0.2,0.4,0.7,1.1,2.0,4.8,100$. \textit{Right}: Brane profiles with 
$L=1,a=1$ and, from outer to inner curve, $c=0.5,1,1.5,...,5$. Note that all 
the turning points lie on a straight line, indicated in dashed black.}
\label{fig::HyperHyper}
\end{figure}

The second, more easily generalisable method is to make use of equation 
\eqref{consistencyAdS} which for the perfect fluid under consideration here 
reads\footnote{This is basically the equation of motion for the perfect fluid. 
Another important equation would be the particle number conservation, which 
in the static case is trivially satisfied as 
\begin{align}
\nabla_i\left(n(\rho(z))u^i\right)= u^i \partial_i n(\rho(z)) + n(\rho(z)) 
\nabla_i u^i=0,
\nonumber
\end{align}
where the last term vanishes by \eqref{nablauvanishes} and the first term 
vanishes by $u^z=0$ and $\partial_t n(\rho(z))=0$. Here, the particle number 
density $n$ is related to the energy density $\rho$ by the equation of state. 
See \cite{Brown:1992kc} for more information on perfect fluids. 
}
\begin{align}
 \rho(z)'=\frac{1+a}{a} \frac{\rho(z)}{z}\ \ \Rightarrow\ \  
\rho(z)=\frac{2c}{\kappa L}\left(\frac{z}{L}\right)^{1+\frac{1}{a}},
 \label{rhosolution}
\end{align}
with $c$ some positive number of our choice. Inserting this into 
\eqref{EOMagain} and solving for the profile $x_{+}(z)$, we 
find (with $y\equiv z/L$),
\begin{align}
x_{+}'(y)&=\frac{acy^{\frac{1}{2}\left(2+\frac{2}{a}\right)}}{\sqrt{
1-a^2c^2y^
{ 2+\frac
{2}{a}}}},
\label{pfODE}
\\
x_{+}(y)&=\frac{a^2c y^{2+\frac{1}{a}} 
{}_{2}F_{1}\left(\frac{1}{2},\frac{1+2 a}{2+2 a};\frac{3+4 a}{2+2 
a};a^2c^2y^{2+\frac{2}{a}}\right)}{1+2 a},
\label{Hypergeo}
\end{align}
with ${}_{2}F_{1}(a,b;c;d)$ the hypergeometric function. It may be shown that 
as $y\rightarrow \left(\frac{1}{a^2c^2}\right)^{\frac{a}{2a+2}}$, the 
derivative diverges while the function itself reaches a finite value. Also, the 
function is real only for $y\leq\left(\frac{1}{a^2c^2}\right)^{\frac{a}{2a+2}}$ 
and below, i.e.~at the critical value and closer to the boundary, for $a\geq0$ 
(i.e.~SEC) and $a\leq-1$. For the intermediate range $-1<a<0$, we do not obtain 
solutions for which the brane actually reaches the boundary, and therefore 
we will ignore this parameter range. While the SEC may easily be violated by 
sensible classical matter (see section \ref{sec::Kondo}), we assume NEC to be 
satisfied (see the discussion at the end of section \ref{sec::ECs2}) and hence 
we will in the following only discuss the solutions for 
$a\geq0$\footnote{In the special $a=-1$, conservation of energy-momentum 
requires $\rho=const.$ and the perfect fluid hence turns into a constant 
tension. We will explicitly allow for a constant tension in section 
\ref{sec::PF+C}.}. See figure \ref{fig::HyperHyper} for plots of the 
embedding function $x_+$ for several values of the parameters. 

At the critical value of $y$ where the derivative of $x_{+}$ diverges, 
the brane should not simply end, but instead be joined with a similar profile 
that symmetrically goes back to the boundary. So the brane starts at the 
boundary and enters the bulk, until at some critical value of $y$ it turns 
around and returns to the boundary in a symmetric fashion. This is interesting, 
since it means that a finite part of the boundary is enclosed by the 
brane and hence by gluing two parts together the spatial boundary of the 
spacetime is compactified. The final result is of the form discussed in section 
\ref{sec::toymodel}, see in particular the right hand sides of figures 
\ref{fig::AdSmatching} and \ref{fig::furtherconsid}.

Will the perfect fluid rather resemble the $\lambda>0$ or the $\lambda<0$ 
case of the discussion in \ref{fig::furtherconsid}? 
As discussed above, for $a\geq0$ the perfect fluid satisfies both WEC and 
SEC. Due to the barrier theorem (see the discussion in section 
\ref{sec::Wall}), we know that extremal curves anchored to one side of the 
brane will not cross it and hence cannot be altered by it. For 
spacelike codimension one extremal surfaces describing entanglement entropy, 
this can be seen easily from figure \ref{fig::HyperHyper}: 
In AdS space the geodesics are half circles, and it can 
be shown that any half-circle anchored to the boundary on one side of the 
brane will never intersect this brane. This is because the brane always 
extends further into the bulk than the largest half-circle that could possibly 
be anchored on one side of the brane, compare $\varUpsilon_2$ and $Q_1$ on the 
left side of figure \ref{fig::drawing1}. In fact, in the limit 
$a\rightarrow+\infty$ the profile \eqref{Hypergeo} asymptotes to the half 
circle drawn as $\varUpsilon_2$ in figure \ref{fig::drawing1}.

\subsection{Perfect fluid in AdS with cosmological constant}
\label{sec::PF+C}

The results obtained above may be generalised to the case where the tension on 
the brane involves a cosmological constant. Generalising \eqref{PFscalarsa}, let 
us now assume
\begin{align}
 S&=\Omega+\rho(a-1),\ \ S_{L/R}=\rho\frac{1+a}{2},
 \label{PFscalarsplusC}
\end{align}
which corresponds to a cosmological constant 
(or equivalently a constant tension) $\Omega$ on the brane\footnote{With this 
ansatz we have $S_{ij}=\Omega/2\gamma_{ij}+S_{matter}$, i.e.~compared to the 
constant tension case investigated in section \ref{sec::toymodel} we have 
$\Omega=-2\lambda$.}. This leads to a situation where the SEC is violated at 
least near the boundary, and may or 
may not become satisfied deeper in the bulk, depending on the energy 
content of the perfect fluid. This situation is interesting in the 
light of the discussion given in section \ref{sec::Wall}: Will the branes 
generally turn around and bend back to the boundary?

$\Omega$ drops out of the equation \eqref{consistencyAdS} when 
differentiating, 
so we end up with the same solution for $\rho$ as before: $\rho(z)=c 
z^{1+\frac{1}{a}}$ where we set $L=1$ and $\kappa=2$ for simplicity. 
By the scalar 
equations of motion \eqref{EOMagain} we now find 
\begin{align}
 \Omega+2a\rho(z)=\mK+2\mK_{L/R}=
 \frac{2 
x_{+}'(z)}{\sqrt{1+x_{+}'(z)^2}}
\end{align}
and integrate with the result
\begin{align}
 x_+'(z)&=\frac{\Omega+2a\rho}{\sqrt{4-(\Omega+2a\rho)^2}}
 \\
 \Rightarrow x_+(z)&=
 \frac{z  }{(1+2 a) \sqrt{4-\Omega^2}}
 \label{PFplusC}
\\
&\times\Bigg[(\Omega +2 a \Omega 
) 
F_1\left(\frac{a}{1+a};\frac{1}{2},\frac{1}{2};\frac{1+2a}{1+a};
-\frac{2 a c z^{1+\frac{1}{a}}}{-2+\Omega },-\frac{2 a c 
z^{1+\frac{1}{a}}}{2+\Omega }\right)
\nonumber
\\
&+2 a^2 c z^{1+\frac{1}{a}} 
F_1\left(\frac{2+a}{1+a};\frac{1}{2},\frac{1}{2};\frac{3+2a}{1+a};
-\frac{2 a c z^{1+\frac{1}{a}}}{-2+\Omega },-\frac{2 a c 
z^{1+\frac{1}{a}}}{2+\Omega }\right)\Bigg]
\nonumber
\end{align}
where $F_1(a;b,c;d;e,f)$ is the Appell hypergeometric function.

As expected, the behaviour is dominated by the cosmological constant near the 
boundary and by the perfect fluid in the bulk, see figure \ref{fig::PFboth} 
(left). This also means that these branes do in fact bend back to the boundary.

Let us briefly mention that in the limit $c\rightarrow0$, the above results 
show us that for a constant tension and no other matter-energy content on the 
brane in Poincar\'e AdS the solution simply reads:
\begin{align}
 x_+(z)=\frac{\Omega  z}{\sqrt{4-\Omega ^2}},
\end{align}
i.e.~the brane is defined by a straight line in \Poincare coordinates as was 
found before \cite{Takayanagi:2011zk,Fujita:2011fp,Nozaki:2012qd}.

\subsection{Perfect fluid in BTZ}
\label{sec::PFinBTZ}

\begin{figure}[htb]
\centering
 \def\svgwidth{0.65\columnwidth}
\executeiffilenewer{PFboth.svg}{PFboth.pdf}%
{inkscape -z -D --file=PFboth.svg %
--export-pdf=PFboth.pdf --export-latex}%
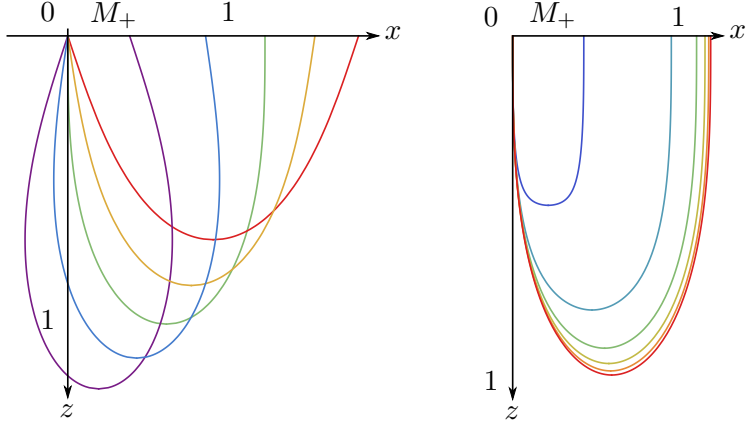%

\caption{\textit{Left:} Embedding functions \eqref{PFplusC} (with additional 
 symmetric branches leading back to the boundary) 
 for $L=1$, $a=c=1$ and, in order of increasing second 
 intersection with the $x$-axis, $\Omega=-1,-0.5,0,0.5,1$.	
\textit{Right:} Embedding functions for the brane with a perfect fluid in a BTZ 
background. Depicted are the cases for $\Omega=0$, $L=a=c=1$ and, in order of 
increasing second intersection with the $x$-axis, $z_H=0.5,1,1.5,...,3$.
}
\label{fig::PFboth}
\end{figure}

Let us repeat the calculation of section \ref{sec::PF+C} in the BTZ background 
\eqref{BTZmetric}, still 
assuming $L=1,\ \kappa=2$ and \eqref{PFscalarsplusC}. The solution to 
\eqref{consistency} 
now reads
\begin{align}
 \rho(z)=c \left(\frac{z^2 z_H^2}{z_H^2-z^2}\right)^{\frac{a+1}{2 a}}.
 \label{rhoinBTZ}
\end{align}
From the equations \eqref{EOMagain} we find again:
\begin{align}
 \Omega+2a\rho(z)=S+2S_{L/R}=\frac{2 z_H x'_+(z)}{\sqrt{(z_H^2-z^2) 
 x_+'(z)^2+z_H^2}}.
\end{align}
Obviously, together with \eqref{rhoinBTZ} this can be solved to give an 
analytic expression for $x'_+(z)$, yet unfortunately this expression cannot be 
integrated to a closed form expression for $x_+(z)$ for general parameters 
$\Omega,a,c,z_H$. In any case, by investigating the limit $z\rightarrow0$ we 
see that, assuming $a>0$, $x'_{+}(z)$ and hence $x_+(z)$ are real for small 
enough $z$ whenever $\Omega^2<4$, as expected. Also assuming $a>0, \Omega^2<4$, 
it may be shown that $x'_{+}(z)$ necessarily diverges for some finite value 
$z_{crit}<z_H$, i.e.~the brane will always bend back to the boundary before 
reaching the event horizon, see figure \ref{fig::PFboth} (right).  

\subsection{Relation to scalar field}
\label{sec::relationtoscalar}

A perfect fluid with an equation of state \eqref{PFeos} may not seem to be
a model of matter that might appear in a `fundamental' holographic 
construction of a given DCFT or BCFT. Nevertheless, we would like to point out 
that at least in the static case the conformal fluid with $a=1$ may be shown to 
be equivalent to a free massless scalar field with action
\begin{align}
  \mL_{matter,\,Q}=-\frac{1}{2}\gamma^{ij}\partial_i \phi\partial_j \phi.
  \label{scalarLagrangian}
\end{align}
Specifically, identifying $\rho\equiv\frac{1}{2}\gamma^{ij}\partial_i 
\phi\partial_j 
\phi=\frac{\gamma^{zz}}{2}\left( \phi'\right)^2$, the equations of motion 
\begin{align}
\gamma'^{11}\phi'- \frac{2}{z}\gamma^{11}\phi'+2\gamma^{11}\phi''=0
\end{align}
following from \eqref{scalarLagrangian} take precisely the form 
\eqref{rhosolution} for a \Poincare background and we also recover 
\eqref{PFscalarsa}. In this way, the solutions presented in the previous 
subsections for $a=1$ can equally be interpreted as solutions for the massless 
free scalar field. 

We conclude with a short summary of what was derived in this section. 
We found explicit analytic solutions for the backreaction in the AdS/BCFT ansatz 
as described in sections \ref{sec::AdSBCFTreview} and \ref{sec::backreaction}.
In particular, we found solutions for the case when the energy-momentum tensor 
on the brane is described by a perfect fluid. Moreover, we also considered the 
case where we add a constant brane tension to a perfect fluid in an ambient AdS 
or BTZ background in 2+1 dimensions. For all of these cases, due to the barrier 
theorem the brane has to bend back to the boundary and hence the spatial 
boundary direction has to be compactified. It will also be interesting to make 
contact between the present results for $a=1$ and the discussion of holographic 
boundary RG flows as considered in \cite{Nakayama:2012ed}.

%%%%%%%%%%%%%%%%%%%%%%%%%%%%%%%%%%%%%%%%%%%%%%%%%%%%%%%%%%%%%%%%%%%%%%%%%%%%%%%%
%%%%%%%%%%%%%%%%%%%%%%%%%%%%%%%%%%%%%%%%%%%%%%%%%%%%%%%%%%%%%%%%%%%%%%%%%%%%%%%%

\section{A holographic model of the Kondo effect}
\label{sec::Kondo}

In this last section, we focus on a particular holographic model, namely 
the holographic Kondo model proposed in \cite{Erdmenger:2013dpa}
(see e.g.~\cite{2001cond.mat..4100K} for a review of the Kondo model in field 
theory). This describes the coupling of an impurity in a 1+1-dimensional CFT, 
i.e.~in a DCFT. Hence we assume that the formalism described in section 
\ref{sec::backreaction} is applicable. In \cite{Erdmenger:2013dpa}, the probe 
limit was considered in which the brane is fixed to be given by a totally 
geodesic brane embedding $x_{\pm}=0$, and the energy-momentum tensor of the 
brane does not backreact to the geometry.

As mentioned before, the probe limit cannot give information about the 
behaviour of entanglement entropy near the defect as it is fixed by the ambient 
geometry, at least for the Ryu-Takayanagi formalism 
\cite{Ryu:2006bv,Ryu:2006ef}. Here, we shed some light on this topic by 
including the backreaction for the dual Kondo system.

In sections \ref{sec::ECs} and \ref{sec::Wall} we saw how critical it is for the
structure of the embedding whether certain energy conditions are satisfied
or violated. In section \ref{sec::PF} we studied a concrete example 
of an energy-momentum tensor on the brane that everywhere satisfies WEC and SEC. 
By the results in table \ref{tab::table1}, SEC implies that initially 
$x'_{+}\geq0$, i.e.~the brane profile $x_+$ initially bends to the right in 
figure \ref{fig::geometry}. Then, by WEC, SEC and the theorem presented in 
section \ref{sec::Wall}, the brane has to reach back to the boundary after a 
finite distance $\Delta x$. This especially means that in 2+1 dimensions, with 
an ansatz relying on matter fields on the brane that satisfy WEC and SEC, it 
will not be possible to find a solution where the brane falls into the event 
horizon of a black hole, as for example in \cite{Magan:2014dwa}. If this was the 
case for all physical models, it would be very restrictive from the point of 
view of model building. 

Hence, in this section we will present a physically motivated ansatz for matter 
fields living on the brane that leads to a SEC violation, and brane embedding 
solutions that do indeed reach into a black hole event horizon. The matter 
content on the brane is given by a scalar field charged under a $U(1)$ gauge 
field, which as well is contrained to the brane, and a $U(1)$ Chern-Simons 
field, which is defined throughout the bulk.

The matter fields on the brane are assumed to be given by the action 
\begin{align}
\mS=-\mN \int \totd^2 x\, \sqrt{-\gamma}\,\left(\frac{1}{4}f_{mn}f^{mn} + 
\gamma^{mn}(D_m \Phi)^{\dagger} (D_n\Phi) +V(\Phi\Phi^{\dagger})\right),
\label{Kondobraneaction}
\end{align}
where $\mN$ is a factor which may be fixed in a top-down string theory 
derivation of the model, see for instance \cite{Erdmenger:2013dpa}. $D$ 
denotes gauge covariant
differentiation w.r.t.~the gauge field $a$ and the Chern-Simons
field $A$, given by
\begin{equation}
 D_m\Phi \equiv \partial_m \Phi + i\,A_m \Phi -i\,a_m \Phi.
\end{equation}
For simplicity, we will henceforth set the bulk Chern-Simons field to zero, 
$A=0$, as it would otherwise contribute to the equations of motion with its own 
junction conditions. 
We will investigate the solutions of the full system 
(including the Chern-Simons field) and their interpretation as a model of the 
Kondo effect in a forthcoming publication \cite{Erdmenger:2015spo}.

\subsection{A gauge field on the brane}
\label{sec::U1fieldonbrane}

Let us first only look on the case of a $U(1)$-field on the brane with 
$\Phi\equiv0$ which resembles the background solution of the Kondo model in the
normal phase. The energy-momentum tensor for the gauge field reads
\begin{equation}
 S^{(a)}_{ij}=-\frac{\mN}{4}f^{mn}f_{mn}\gamma_{ij}+\mN\gamma^{mn}f_{mi}f_{nj}
\label{eq:energystressYM}
\end{equation}
which can be decomposed into a trace part
\begin{equation}
S^{(a)}=\frac{\mN}{2}f^{mn}f_{mn}
\end{equation}
and a traceless part given by
\begin{align}
S^{(a)}_{ij}-\frac{S^{(a)}}{2}\gamma_{ij}=-\frac{\mN}{2}f^{mn}f_{mn}\gamma_{ij}+\mN\gamma^{
mn} f_
{
mi}f_{nj}.
\label{notracepart}
\end{align}
The equations of motion in the absence of currents
\begin{align}
 \partial_m\left(\sqrt{-\gamma}\gamma^{mp}\gamma^{nq}f_{pq}\right)=0
 \label{eq:EOMgaugeWOcurrent}
\end{align}
imply due to antisymmetry $\sqrt{-\gamma}f^{01}=:C$, with $C$ constant. 
This is the electric flux due to the gauge field.

We will now assume a static configuration which means, as we saw in section 
\ref{sec::staticcase}, that the metric takes the form 
\eqref{abmetric}\footnote{From here on, our arguments will also be valid in 
the case $C\neq const$ as applies for a current present in 
\eqref{eq:EOMgaugeWOcurrent}.}. 
As the tensor $f_{mn}$ is antisymmetric and hence has only 
one component that can be specified in 1+1-dimensions, it is straightforward 
to show $f^{mn}f_{mn}=-2C^2$ and $\gamma^{mn}f_{mi}f_{nj}=-\gamma_{ij}C^2$. 
This means that the traceless part \eqref{notracepart} vanishes identically, 
and 
the energy-momentum tensor takes the form
\begin{align}
 S^{(a)}_{ij}=-\frac{\mN}{2}\gamma_{ij}C^2
\end{align}
which corresponds to a (WEC satisfying, SEC violating) constant tension
$\lambda=\mN C^2 /2$. 

\begin{figure}[ht]
 \centering
 \def\svgwidth{0.8\columnwidth}
\executeiffilenewer{NormalFlow.svg}{NormalFlow.pdf}%
{inkscape -z -D --file=NormalFlow.svg %
--export-pdf=NormalFlow.pdf --export-latex}%
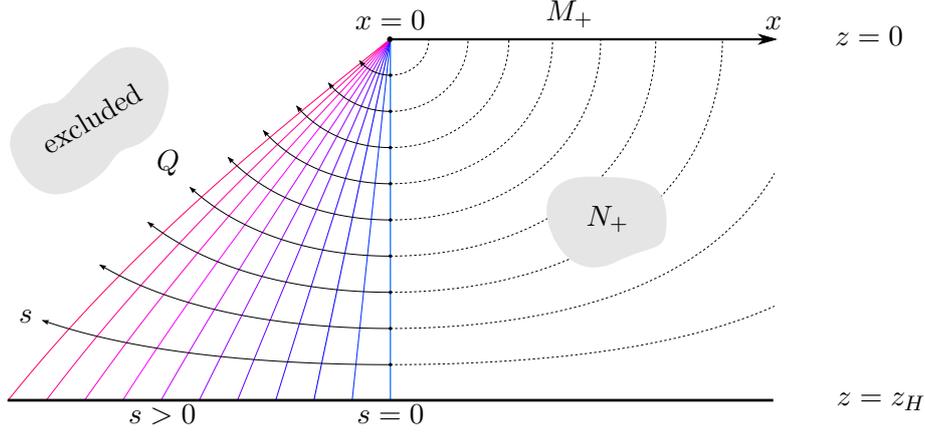%

\caption{Construction of constant brane tension solutions in a BTZ geometry 
using a geodesic normal flow. 
We start from the trivial, totally geodesic embedding $x_+ =0$ (for which $s=0$) 
and follow the flow generated by geodesics normal to the trivial embedding and
pointing outward of $N_+$. The region to the left of the family of embeddings 
is excluded from $N_+$. 
The dashed lines to the right hand side denote the continuation of the 
generating geodesics.
}
\label{fig::normalFlow}
\end{figure}

Motivated by the results in \cite{Azeyanagi:2007qj} (see our section 
\ref{sec::toymodel}) in which the authors obtained that the final embedding can 
be reached via a geodesic normal flow, we now investigate whether this can be 
generalised to other spacetimes as well. The construction is then as follows, 
see figure \ref{fig::normalFlow}: Starting from the trivial embedding $x_+=0$, 
we follow the geodesics starting normal to the hypersurface for a certain 
arclength $s$, depicted by the solid black lines. The arclength $s$ can be 
matched to the brane tension $\lambda$ using 
\eqref{eq::TakayanagiGeodesicLengthAndBraneTension}, in
which we have to identify $y_{\ast}$ with $s$ such that
\begin{equation}
 \tanh(s/L)=\frac{L}{2}\kappa\lambda=\frac{L}{4}\kappa \mN C^2 .
 \label{eq::GeodesicLengthAndElectricFlux}
\end{equation}
This relationship remains valid for our case with a BTZ background and even 
more general spacetimes, see appendix \ref{app::normalflow}. 
Note that \eqref{eq::GeodesicLengthAndElectricFlux} yields a bound on the energy 
momentum tensor on the brane. In particular, for our matter content it reads
\begin{equation}
 |L \kappa \mN C^2| < 4.
 \label{BoundMatter}
\end{equation}
For the limiting case, $s\rightarrow \infty$, the embedding of $Q$ is given
by the conformal boundary and the construction considered is no longer 
applicable.

The family of embeddings $X_s:Q\hookrightarrow N_+$ is 
explicitly given by

\begin{equation}
X_{s}(t,z_{b}) = (t,x_+(z_b),z(z_b)) = 
\left( 
t , 
- z_H\, \arctanh\left(\frac{z_{b}}{z_H}\tanh(\frac{s}{L})\right) , 
\frac{z_{b}}{\sqrt{1 + f(z_{b})\sinh^2(\frac{s}{L})}} 
\right)
 \label{normalcongruence3dim}
\end{equation}

with $(t,z_b)$ the point at which the geodesics start on the trivial embedding.
The induced metric and the extrinsic curvature are computed to be
\begin{equation}
 \gamma_s = f_{s}^{\,\ast} g 
 = \left(\frac{L\cosh(s/L)}{z_{b}}\right)^2 \left(-f(z_{b})\,
   \totd t^2+f^{-1}(z_{b})\,\totd z_{b}^2\right)
\end{equation}
\begin{equation}
 K_s = 
 \frac{-L\sinh (s/L) \cosh (s/L)}{z_{b}^2}\left(-f(z_{b})\,
 \totd t^2 + f^{-1}(z_{b}) \,\totd z_{b}^2\right).
\end{equation}

Note that the normal flow changes the induced metric only by a conformal
factor. Furthermore, the extrinsic curvature is proportional to the induced
metric
\begin{equation}
K_s = \frac{-\tanh(s/L)}{L} \gamma_s
\label{PropotionalityExtrCurvToIndMetricBTZ}
\end{equation}
and for $s=0$ we recover the trivial embedding. This proportionality is 
necessary to satisfy the Isreal junction conditions for constant tensions, see
also appendix \ref{app::normalflow}.

To obtain explicit solutions in a BTZ background, we can choose the gauge
$a_z=0$ and integrate $\partial_z a_t = \sqrt{-\gamma} C$ with boundary
condition $a_t(z_H)=0$ for regularity. 
Moreover we may solve \eqref{normalcongruence3dim} for $x_+$ as a 
function of $z$ for constant $s$. We obtain

\begin{equation}
 \begin{aligned}
  a_t 	
   = \int\limits^z_{z_H} \sqrt{-\gamma} \,C
  &= \frac{C L^2}{z_H} \cosh (s/L) 
     \left(\cosh(s/L)-\sqrt{(z_H/z)^2+\sinh^2 (s/L)}\right)\\
  x_+(z)
  &=-z_H \,\arctanh\left(\frac{\sinh(s/L)}{\sqrt{(z_H/z)^2+\sinh^2(s/L)}}\right)
\end{aligned}
 \label{backgroundSolutionGaugeAndEmbedding}
\end{equation}
where the arclength $s$ and the electric flux $C$ still have to be matched 
according to
\eqref{eq::GeodesicLengthAndElectricFlux} in order to satisfy the equations of 
motion.

\subsection{A gauge field with non-trivial scalar on the brane}

Now that we found the background solutions
\eqref{backgroundSolutionGaugeAndEmbedding}, we allow the scalar field in 
\eqref{Kondobraneaction} to be non-vanishing. 
In the holographic Kondo model, this amounts to 
an RG flow triggered by a marginally relevant operator, and to
a phase transition to the condensed 
phase 
in the large $N$ limit \cite{Erdmenger:2013dpa}.
This corresponds to the formation of the Kondo screening cloud.
Upon turning on the scalar field we have to add the scalar sector
of the energy-momentum tensor to \eqref{eq:energystressYM}, which reads
\begin{align}
 S^{(\Phi)}_{ij}=2\mN\left[(D_{(i} \Phi)^{\dagger} D_{j)} \Phi 
-\frac{1}{2}\gamma_{ij}\left(|D\Phi|^2+V(\Phi^{\dagger}\Phi)\right)\right].
\end{align}
The total energy-momentum tensor is hence given by
\begin{equation}
 S^{(tot)}_{ij} = S^{(a)}_{ij} + S^{(\Phi)}_{ij} .
\end{equation}
For the static case in two dimensions the scalar part may be decomposed as
\begin{align}
  S^{(\Phi)}=-2\mN V(\Phi^{\dagger}\Phi),\ \ 
S^{(\Phi)}_{L/R}=\frac{\mN }{2}\tilde{\gamma}^{ij}(D_i\Phi)^{\dagger} D_j\Phi,
\label{decompositionScalar}
\end{align}
$\tilde{\gamma}_{ij}$ being defined in \eqref{gammatilde}. As the latter is 
manifestly positive, we see that the scalar field always satisfies NEC. 

\begin{figure}[htb]
\centering
\def\svgwidth{0.9\columnwidth}
\executeiffilenewer{backreaction_grid_emb_Rotated.svg}{backreaction_grid_emb_Rotated.pdf}%
{inkscape -z -D --file=backreaction_grid_emb_Rotated.svg %
--export-pdf=backreaction_grid_emb_Rotated.pdf --export-latex}%
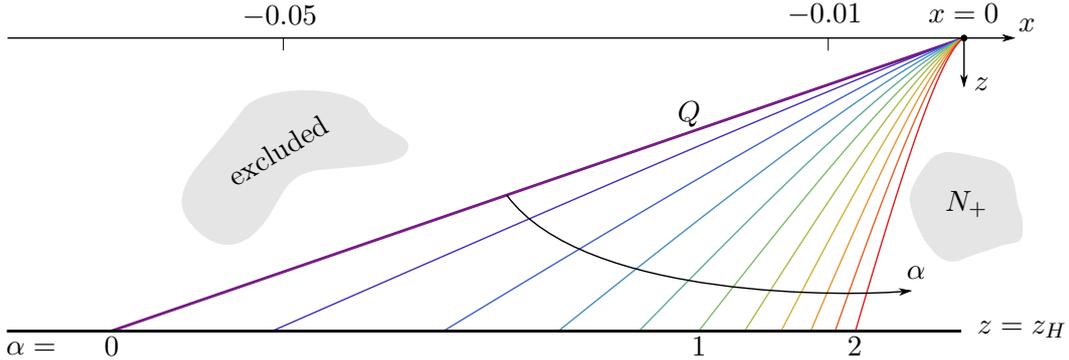%

\caption{Numerical solutions for the embedding function $x_+(z)$ for different
ratios of $\mu/T$. 
The region between the embedding profile and the $x$-axis is
excluded from the spacetime $N_+$ as indicated in the plot.
The 
violet curve (thick) is given by a vanishing scalar field $\alpha=0$ and hence 
\eqref{backgroundSolutionGaugeAndEmbedding}, where the arclength $s$ has been 
tuned to be consistent with the gauge field background configuration.
The other embeddings are given by numerical solutions for 
$\alpha=0,0.2,0.4,\ldots,2$ where we set $L=1$, $z_H=1$, $C=1/2$ and 
$\kappa=1$.
Increasing $\alpha$ is equivalent to increasing the ratio of $\mu/T$, hence the
overall volume of the spacetime decreases as the system undergoes a phase 
transition from the normal to the condensed phase in the large $N$ limit.
The condensed phase corresponds to the formation of the Kondo screening cloud.}
\label{fig::numericalEmb}
\end{figure}

To apply the results of section \ref{sec::Wall}, we need to make statements
about the violation of WEC and/or SEC as well. For the holographic Kondo model,
we find that in contrast to the perfect fluid solutions discussed in section
\ref{sec::PF}, the SEC is violated everywhere on the brane. This can easily be 
seen from the numerical solutions in figure \ref{fig::numericalEmb}, since here
$x'<0$ between the conformal boundary and the event horizon. Due to the results
of table \ref{tab::table1}, this corresponds to SEC violation.

The equations of motion for the scalar $\Phi$, the gauge field $a_t$ 
(we gauge $a_z=0$) and the embedding scalar $x_+$ read
\begin{align}
 \gamma^{\alpha\beta} \mathfrak{D}_{\alpha} \mathfrak{D}_{\beta} \Phi
 -M^2 \Phi &= 0, \label{eq:EOMscalar}\\
 \partial_z\sqrt{-\gamma}f^{zt} + J^t 
 &=0, \label{eq:EOMgauge}\\
 \mK_{zz} - \frac{\kappa}{2} S^{(tot)}_{zz}
 &=0, \label{eq:EOMemb}
\end{align}
where we defined the gauge covariant derivative 
$\mathfrak{D} = \nabla + i a$ with $\nabla$ the induced connection
on the brane and the conserved current
$J^{\mu} = - 2 \sqrt{-\gamma} \gamma^{\mu\nu} (a_{\nu})\Phi^2$.
Regularity at the event horizon requires
\begin{equation}
 \Phi'(z_H) = -\frac{L^2\,M^2 \Phi(z_H)}{2 z_H}, \qquad a_t(z_H)=0, \qquad 
x_+'(z_H) = \kappa\,\frac{2L^4 M^2  \Phi(z_H)^2-z_H^{\,4} \,a_t'(z_H)^2}{4 L^3},
\end{equation}
which fixes half of the integration constants. Since the embedding enters the 
EOMs only via its first and second derivative, its remaining integration 
constant can easily be fixed by requiring $x_{\pm}(0)=0$. 
This is resonable 
since the ambient spacetime is invariant under translations in $x$-direction.

We start by treating the scalar field as a probe w.r.t.~the background 
solution \eqref{backgroundSolutionGaugeAndEmbedding}.
The leading order behaviour of the gauge field reads
\begin{equation}
  a_t \sim \frac{- C L^2 \cosh(s/L)}{z} 
  + \frac{C L^2 \cosh^2(s/L)}{z_H} + \ldots =: Q/z + \mu_c + \ldots
\end{equation}
where we defined $Q = - C L^2 \cosh(s/L)$ and $\mu_c = C L^2 \cosh^2(s/L)/z_H$, 
the latter being identified with the chemical potential for the $U(1)$ charge.
If the scalar is nonzero, we have $\mu>\mu_c$.
Furthermore, from \eqref{eq::GeodesicLengthAndElectricFlux}
we obtain $L\kappa C^2 = 4 \tanh(s/L)$ and hence a restriction on the 
magnitude of $C$ as mentioned above.

Turning on the scalar near the boundary, we find that there are
two asymptotic solutions which read
\begin{equation}
 \Phi \sim z^p, \qquad p = 
\frac{1}{2}\left(1\pm\sqrt{1+4(L^2M^2-Q^2)\cosh^2(s/L)}\right).
\end{equation}
To obtain the correct operator dimensions for mapping our bottom-up model to the 
Kondo
effect, just as in \cite{Erdmenger:2013dpa} we need $p=1/2$. This means that our
scalar field saturates the Breitenlohner-Freedman bound and 
we have to adjust the mass of the scalar to be given by 

\begin{equation}
 M^2 = \frac{4 Q^2 \cosh ^2\left(s/L\right)-1}{4 L^2 \,\cosh^2 
\left(s/L\right)} = \left(\frac{Q}{L}\right)^2 - (4 L^2 \,\cosh^2 
\left(s/L\right))^{-1}.
\end{equation}
The asymptotic behaviour of the scalar is then given by
\begin{equation}
 \Phi(z) \sim \alpha \sqrt{z} \log(z \Lambda) + \beta \sqrt{z} + \ldots
\end{equation}
where we introduced an arbitrary energy scale $\Lambda$ to define the logarithm. 
It was shown in \cite{Erdmenger:2013dpa} that switching on the scalar
triggers the running of the Kondo coupling. 
The dual operator is similar to the double trace operator considered 
in \cite{Witten:2001ua}. 
Thus the Kondo coupling is given by $\kappa = \alpha/\beta$.

The leading order coefficient $\alpha$ is 
proportional to the vacuum expectation value of the dual operator on the field
theory side,
\begin{equation}
 \alpha \sim \langle \mathcal{O} \rangle,
\end{equation}
such that a non-vanishing scalar field leads to the 
condensation of the Kondo singlet in the field theory.
For further details we refer the reader to \cite{Erdmenger:2013dpa}.

Now fixing $L$, $z_H$, $\kappa$ and $C$ we solve for $Q$, $M^2$ 
and $s$ by using the formulae above. We may turn on the scalar field by either 
increasing the chemical potential $\mu$ or the leading order coefficient 
$\alpha$ of the scalar. 
Both approaches are equivalent for our considerations.
The phase transition occurs at $\mu = \mu_c$,
which in the limit $\kappa\rightarrow 0$ (and $s\rightarrow 0$) yields 
the value for the probe solution $\mu = -Q$.
Note that the critical value for the chemical potential diverges if 
we saturate the bound \eqref{BoundMatter} discussed above.

To solve the system of ODEs \eqref{eq:EOMscalar}-\eqref{eq:EOMemb}, we apply 
the finite difference method. 
We fix $L=1$, $\kappa=1$, $C=1/2$ and $z_H=1$ for concreteness.
To have control on the steep behaviour of the 
fields near the boundary, we subdivided the radial direction $z\in (0,1)$ into 
two subregions $z_L = (0,0.1)$ and $z_R=(0.1,1)$. 
In $z_L$, the ODEs were discretised
on an evenly spaced grid with 300 points while in $z_R$, we applied a Chebyshev
grid with 50 points.

Results for the embedding scalar $x_+(z)$ for different values of 
$\alpha \sim \langle \mathcal{O} \rangle$
are shown in figure \ref{fig::numericalEmb}. 
We see that turning on the 
scalar field decreases the overall volume of the spacetime, 
in contrast to what happens if we increase the magnitude of the electric 
flux $C$.
At the boundary, the behaviour of the embedding does not change as the scalar 
field condenses.
Its derivative at the boundary $x_+'(0)$ is fixed by the asymptotics of the 
gauge field and thus boundary data, 
i.e.~the brane embedding approaches the conformal boundary 
tangential to the background embedding.

\subsection{Summary}

To summarise, we have found analytic solutions for
the brane embedding and the matter field configurations in equilibrium for the 
case of vanishing scalar field on the brane. In that case, the energy-momentum
tensor is given by a constant brane tension and we may construct the embedding
by following a geodesic normal flow.

For constant brane tension solutions, the entanglement entropy of symmetric 
patches around the defect will only differ from that of the trivial solution by 
a constant offset since due to the refraction conditions at the brane\footnote{
The minimisation problem for spacelike curves yields that the geodesics 
connecting boundary points located symmetric around the defect have to start
normal to the brane embedding.}, the generating geodesics are part of 
the extremal surfaces (dashed lines in figure \ref{fig::normalFlow}) used in 
the Ryu-Takayanagi formalism.

Turning on the scalar field decrases the volume of $N_+$ and the energy 
momentum tensor is not given by a constant tension anymore. This can be seen 
from \eqref{decompositionScalar}, which implies $S_{L/R}\neq0$ for a 
non-vanishing scalar. Hence, the corresponding brane embedding cannot be 
constructed by a geodesic flow, which means especially that the entanglement 
entropy of symmetric patches around the defect will be a function of the 
boundary separation. Renormalising this function by subtracting the normal phase 
solution with $\Phi=0$ will yield a non-constant function from which we can 
extract essential information about the Kondo screening cloud and the associated 
defect entropy. This will be investigated in more detail in \cite{Erdmenger:2015spo}.

The Kondo model realises a boundary RG flow in agreement with the holographic
$g$-theorem. This theorem states that the boundary entropy $S_{bnd}$ decreases
along the RG flow \cite{Affleck:1991tk,Friedan:2003yc}. 
From the holographic point of view, this can be derived
by requiring the NEC in the bulk and on the 
brane \cite{Takayanagi:2011zk,Fujita:2011fp}.
In \cite{Takayanagi:2011zk,Fujita:2011fp}, a $g$-function of the form
\begin{equation} 
 \log(g) \sim - \arcsinh\left(\frac{x_+(z)}{z}\right)
 \label{gfun}
\end{equation}
was suggested to give the correct answer for the boundary entropy in AdS.
We may apply this function to our example as well, since due to the SEC
violation, NEC implies $x''>0$ in a BTZ background 
(see table \ref{tab::table1}). This is essential to prove that this function 
decreases monotonically along the RG flow.

%%%%%%%%%%%%%%%%%%%%%%%%%%%%%%%%%%%%%%%%%%%%%%%%%%%%%%%%%%%%%%%%%%%%%%%%%%%%%%%%
%%%%%%%%%%%%%%%%%%%%%%%%%%%%%%%%%%%%%%%%%%%%%%%%%%%%%%%%%%%%%%%%%%%%%%%%%%%%%%%%

\section{Conclusions and outlook}
\label{sec::conc}

Motivated by a recent holographic Kondo model
\cite{Erdmenger:2013dpa}, we have studied gravity configurations with
matter fields restricted to infinitely thin co-dimension one
hypersurfaces, i.e.~branes. These configurations are described by
equations of motion that involve the extrinsic curvature of these 
hypersurfaces, as used in the AdS/BCFT correspondence.

In section \ref{sec::backreaction}, we considered the gravity dual of
a DCFT, where the equations of motion of the brane
$Q$  are the Israel junction conditions. This setup may be 
considered to describe an infinitely thin brane or to describe an approximation 
to a finitely thin matter configuration. 
Moreover, we studied the ensuing equations of motion for 
non-trivial matter fields living on $Q$, both by general arguments and by 
considering concrete examples. 

Section \ref{sec::ECs} was devoted to decomposing the Israel 
junction conditions into its trace and off-trace parts, and to a study of the 
restrictions that energy conditions may impose on the geometry. 
In section \ref{sec::Wall}, we related these energy conditions to a geometrical 
theorem, the `barrier theorem', recently obtained by Engelhardt and Wall. 
Moreover, we deduced qualitative statements on the form that 
a brane $Q$ may take, depending on whether certain energy conditions are 
satisfied or violated by the matter fields living on it. We found that
if the weak and strong energy conditions are satisfied, the brane has
to bend back to the boundary. If just the null energy condition is
satisfied, the brane may reach infinitely far into the radial
direction of the dual space.

In section \ref{sec::toymodel}, we discussed how the
well-known toy model of a brane with constant tension arises in our
context. Then, in section \ref{sec::PF},
we found explicit results for the backreacted configurations 
corresponding to branes whose matter content is given by a perfect
fluid, or - equivalently, as we showed - by  
massless free scalar fields restricted to them. 
Section \ref{sec::Kondo} was then devoted to the numerical study of a brane 
with a matter content that  appears in the holographic Kondo model 
of \cite{Erdmenger:2013dpa}. 
Due to its 
violation of the strong energy condition, it reaches all the way to the
BTZ horizon in the radial direction and thus shows a qualitatively different 
behaviour than the bending solutions studied in section \ref{sec::PF}.

Finally, let us give an outline to the three appendices included below.
We comment on the generalisation of the findings of section \ref{sec::Wall} to 
special geometries in higher dimensions in appendix \ref{sec::Wall2}. 
In appendix \ref{app::normalflow}, we justify the geodesic normal flow 
construction of constant brane tension solutions in sections \ref{sec::toymodel} 
and \ref{sec::Kondo}, and state under which conditions this may be extended to 
more general setups.

We conclude this work with an outlook on interesting topics that in our 
eyes warrant further investigation:

\textbf{The Kondo model:} The first issue is of course the holographic model of 
the Kondo effect \cite{Erdmenger:2013dpa}. We presented preliminary results for 
the brane embedding in the backreacted case in section \ref{sec::Kondo}. 
However, further important questions were left unanswered, 
for example calculating the entanglement entropy for these solutions. 
Moreover, the behaviour of the field theory dual to the backreacted solution 
remains undetermined.
Also, the model 
proposed in \cite{Erdmenger:2013dpa} contains a bulk Chern-Simons-field that we 
set to zero in section \ref{sec::Kondo}. In general however, this field 
requires its own junction conditions. Interestingly, AdS/BCFT like setups with 
Chern-Simons fields in the bulk were studied before in 
\cite{Fujita:2012fp,Melnikov:2012tb}. Building upon the results of this work, 
we intend to revisit all of these questions in a future publication 
\cite{Erdmenger:2015spo}.

\textbf{Exact solutions:} In section 
\ref{sec::PF} we presented analytical solutions for the case where the matter 
living on the brane is a perfect fluid with simple equation of state $p=a\rho$. 
While this may not be a very realistic model in general, we pointed out in 
section \ref{sec::relationtoscalar} that for the choice $a=1$ it describes a 
massless free scalar on the brane.
A further interesting question will be to evaluate the on-shell action and 
examining the thermodynamic behaviour of these solutions.

\textbf{Quantum Hall effect:} Furthermore, we would like to shortly comment 
on the AdS/BCFT model of the 
quantum Hall effect presented in \cite{Melnikov:2012tb}\footnote{An earlier 
study of the Hall effect using an AdS/BCFT ansatz is \cite{Fujita:2012fp}.}. 
There (see especially figure 2(d) of \cite{Melnikov:2012tb}), it was argued that 
a realistic model of the quantum Hall effect would likely include a brane that 
is anchored to the boundary twice similar to $Q_1$ in figure 
\ref{fig::drawing1}. From our results in section \ref{sec::Wall}, it is now 
clear that in $2+1$-dimensions the brane can be forced to show such a behaviour 
by choosing its matter content to satisfy both weak and strong energy condition 
everywhere. 
However, the Hall model of \cite{Melnikov:2012tb} naturally lives in 
$3+1$-bulk dimensions, so it would be desirable to generalise the findings from 
section \ref{sec::Wall} to higher dimensions. We will shortly comment on this 
in appendix \ref{sec::Wall2}.  

\textbf{Holographic bilayers:} We note that bending brane configurations similar
to those we considered in section \ref{sec::PF} also appear in recent 
holographic studies of bilayers 
\cite{Evans:2013jma,Evans:2014mva,Grignani:2014vaa}.
Those studies were performed in top-down probe models.
It will be interesting to apply our backreaction results to these bilayer 
models as well.

\textbf{Transitions from SEC to SEC violation:} As we saw in sections 
\ref{sec::Wall}, \ref{sec::toymodel}, \ref{sec::PF} and \ref{sec::Kondo}, the 
strong energy condition (SEC) and whether it is satisfied or violated has 
tremendous qualitative influence on the dual theory. 
So suppose it were possible 
to tailor a sensible holographic model, in which by varying some physical order 
parameter $\sigma$ (which may e.g.~be a temperature, a charge or something 
else) we obtain a transition from a regime where SEC (but possibly not the 
weak energy condition (WEC)) is satisfied everywhere on the brane to a regime 
where the SEC (but likely not the WEC) are violated everywhere. 
For example this might be possible if the equations of state of some system 
depends on a tunable parameter. 
We have seen in section \ref{sec::toymodel} by varying $\lambda$ 
from positive to negative values that this transition, while nice and steady in 
the bulk, would manifest itself in a very sudden and profound way in the values 
of entanglement entropy of certain boundary intervals, see figure 
\ref{fig::furtherconsid}. 
Specifically, this transition would set in in larger intervals $[AB]$ first,
and then gradually in smaller and smaller 
intervals as the hypersurface $Q$ moves closer to the boundary. 
Yet, from our discussion in section \ref{sec::Wall}, we should point 
out that such results would likely not generalise well to higher dimensions.

%%%%%%%%%%%%%%%%%%%%%%%%%%%%%%%%%%%%%%%%%%%%%%%%%%%%%%%%%%%%%%%%%%%%%%%%%%%%%%%%
%%%%%%%%%%%%%%%%%%%%%%%%%%%%%%%%%%%%%%%%%%%%%%%%%%%%%%%%%%%%%%%%%%%%%%%%%%%%%%%%

\section*{Acknowledgements}

We would like to thank 
Martin Ammon, 
Daniel Fern\'andez, 
Daniela Herschmann, 
Carlos Hoyos, 
Abhiram Kidambi, 
Matthew Lippert, 
Andy O'Bannon, 
Thore Posske, 
Dennis Schimmel, 
Charlotte Sleight, 
Migael Strydom, 
Tadashi Takayanagi 
and Jackson Wu 
for countless helpful discussions.

%%%%%%%%%%%%%%%%%%%%%%%%%%%%%%%%%%%%%%%%%%%%%%%%%%%%%%%%%%%%%%%%%%%%%%%%%%%%%%%%
%%%%%%%%%%%%%%%%%%%%%%%%%%%%%%%%%%%%%%%%%%%%%%%%%%%%%%%%%%%%%%%%%%%%%%%%%%%%%%%%

\appendix

%%%%%%%%%%%%%%%%%%%%%%%%%%%%%%%%%%%%%%%%%%%%%%%%%%%%%%%%%%%%%%%%%%%%%%%%%%%%%%%%
%%%%%%%%%%%%%%%%%%%%%%%%%%%%%%%%%%%%%%%%%%%%%%%%%%%%%%%%%%%%%%%%%%%%%%%%%%%%%%%%

\section{Implications of energy conditions in higher dimensions}
\label{sec::Wall2}

From our discussion in section \ref{sec::conc}, we note that it will 
interesting to generalise the findings of section \ref{sec::Wall} to higher 
dimensions. 
This is involved in general. For illustration, let us consider some examples.

It is indeed easy to construct examples showing that when the brane has more 
than 1+1 dimensions, WEC and SEC together are not enough to imply the 
assumption made in the barrier theorem. Obviously, the analogue of equation 
\eqref{Wall} will always be related to some SEC-like energy condition for 
timelike vectors. The problem is constraining the spacelike vector fields $v^i$ 
by some condition on $S_{ij}$ that can easily be checked for a given matter 
content. 
We have some numerical evidence based on radom matrices 
that in three dimensions on the brane, 
the SEC together with the dominant energy condition (DEC, 
\cite{Visser:1999de,Curiel:2014zba}) and the determinant energy condition 
(DetEC, \cite{Curiel:2014zba,Martin-Moruno:2013wfa}) will be sufficient to 
imply $K_{ij}v^{i}v^{j}\leq0$ for any $v^i$.

\begin{figure}[htbp]
\begin{center}
 \includegraphics[width=0.95\textwidth]{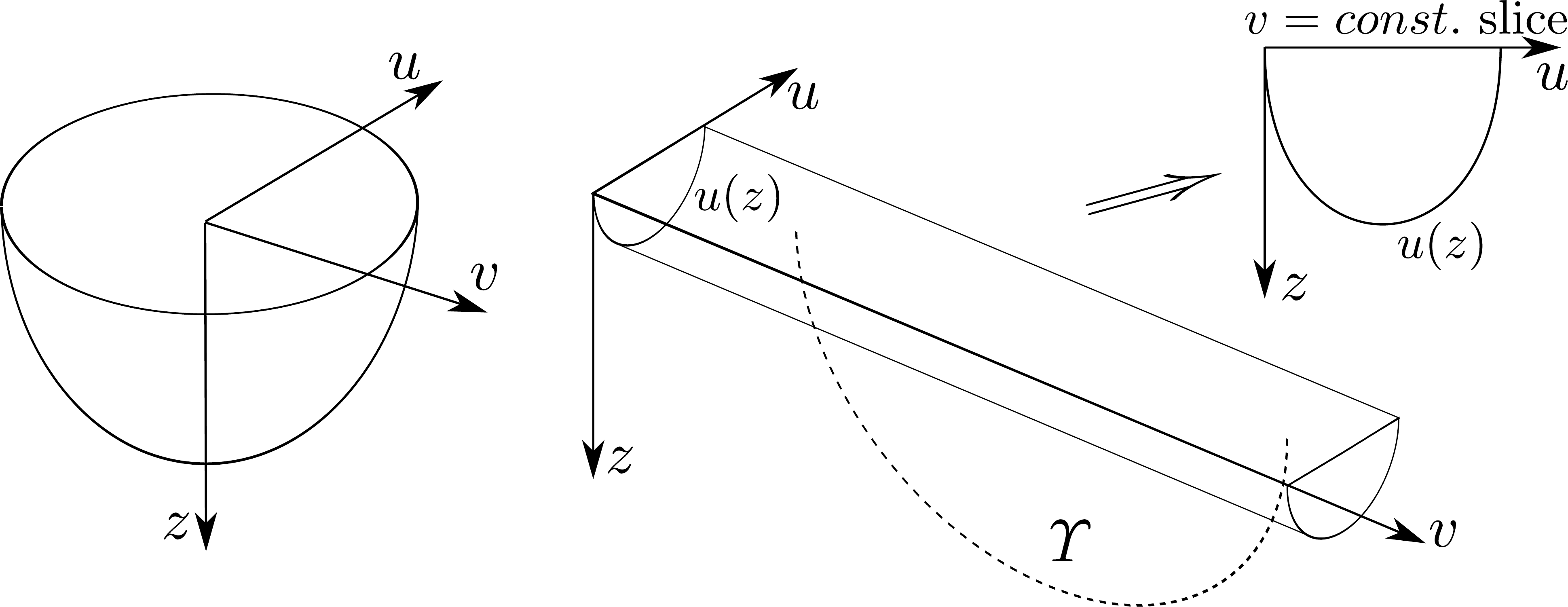}
\caption{Two interesting geometries for a 2+1-dimensional hypersurface $Q$ in a 
3+1-dimensional bulk. The time direction is suppressed, $z$ is the radial AdS 
coordinate, and $u,v$ are boundary coordinates. \textit{Left}: Spherical 
boundary region. \textit{Right}: Strip like boundary region. As indicated, this 
can be mapped to the lower-dimensional case.}
\label{fig::strip}
\end{center}
\end{figure}

One particular problem of the generalisation to higher dimensions is that it may 
lead to geometries as shown on the left hand side of figure \ref{fig::strip}.
However, most applications require a strip geometry.

Let us investigate such a strip geometry where the boundary region is infinitely 
extended, see the right of figure \ref{fig::strip}. 
This geometry may be interesting for the 
holographic study of the Hall effect, see \cite{Melnikov:2012tb} and our short 
discussion in section \ref{sec::conc}. 
Can we use arguments 
similar to section \ref{sec::Wall} to determine some condition on the 
energy-momentum tensor that will force the brane to bend over and come back to 
the boundary as depicted in the figure? 
As can easily be seen, geodesics 
$\varUpsilon$ along the direction of the strip would sooner or later cross it if 
their endpoints are taken far enough apart. 
This requires to enforce the extremal surface barrier property only in a certain 
direction. 
For simplicity, let us consider an embedding in \Poincare-AdS with $L=1$ and 
$\kappa=2$.

For a brane as depicted on the right side of figure \ref{fig::strip} and 
defined by an embedding function $u(z)$, the extrinsic curvature reads 
(coordinates $x^0=t, x^1=z, x^2=v$)
\begin{align}
 K_{\alpha\beta}=
 \left(
\begin{array}{ccc}
 -\frac{u'(z)}{z ^2 \sqrt{1+u'(z)^2}} & 0 & 0 \\
 0 & \frac{u'(z)+u'(z)^3-z  u''(z)}{z ^2 
\sqrt{1+u'(z)^2}} & 0 \\
 0 & 0 & \frac{u'(z)}{z ^2 \sqrt{1+u'(z)^2}} \\
\end{array}
\right).
\end{align}
The first diagonal entries are the extrinsic curvature of the profile curve 
embedded in $2+1$-dimensional \Poincare space, so it is possible to map the 
system to the lower-dimensional case of section \ref{sec::Wall}.
Apart from the trivial case $u'(z)=0$, the brane will be forced to return to 
the boundary when $u'(z)\geq0$ and $u'(z)+u'(z)^3-z  u''(z)\leq0$, see section 
\ref{sec::Wall} and compare to the left hand side of figure 
\ref{fig::drawing1}. 
Interestingly, this will necessarily imply $K_{yy}\geq0$.
As expected, the brane will not be an extremal surface barrier for spacelike 
extremal surfaces not contained in a $v=const.$ slice. 
This applies e.g.~to the geodesic $\varUpsilon$ in figure \ref{fig::strip}. 

The energy-momentum tensor on the brane is given by
$S_{\alpha\beta}=-K_{\alpha\beta}+\gamma_{\alpha\beta}K$ and hence
\begin{align}
 S_{\alpha\beta}=
 \left(
\begin{array}{ccc}
 -\frac{2 \left(u'+u'^3\right)-z  
u''}{z ^2 \left(1+u'^2\right)^{3/2}} & 0 & 0 
\\
 0 & \frac{2 u' \sqrt{1+u'^2}}{z ^2} & 0 \\
 0 & 0 & -\frac{-2 \left(u'+u'^3\right)+z  
u''}{z ^2 \left(1+u'^2\right)^{3/2}} \\
\end{array}
\right).
\end{align}
In the following we will only use vectors with vanishing $v$-component, 
i.e.~vectors that are contained in a $v=const.$ slice. First of all, NEC implies 
$u''\geq 0$, WEC implies $2u'+2u'^3-zu''\leq0$ and together, these imply
\begin{align}
 u'+u'^3-zu''\leq0
\end{align}
which is the one component in $K$. Let us project out the $v$-direction, so 
that we have 
\begin{align}
  K^{\bot}_{ij}&=
 \left(
\begin{array}{cc}
 -\frac{u'}{z ^2 \sqrt{1+u'^2}} & 0  \\
 0 & \frac{u'+u'^3-z  u''}{z ^2 
\sqrt{1+u'^2}} \\
\end{array}
\right),
\\
 S^{\bot}_{ij}&=
 \left(
\begin{array}{cc}
 -\frac{2 \left(u'+u'^3\right)-z  
u''}{z ^2 \left(1+u'^2\right)^{3/2}} & 0  
\\
 0 & \frac{2 u' \sqrt{1+u'^2}}{z ^2} \\
\end{array}
\right),
\\
\gamma^{\bot}_{ij}&=
 \left(
\begin{array}{cc}
 -\frac{1}{z ^2} & 0  \\
 0 & \frac{1+u'^2}{z ^2}  \\
\end{array}
\right).
\end{align}
Then the replacement of the SEC
\begin{align}
 \left(S^{\bot}_{ij}-\gamma^{\bot}_{ij}S^{\bot}\right)v^iv^j\geq0,
\end{align}
with $S^{\bot}=\gamma^{\bot ij}S^{\bot}_{ij}$ implies $u'\geq0$ which together 
with the NEC ensures that 
$K^{\bot}_{ij}$ (but not $K_{\alpha\beta}$) is negative definite. By choosing 
matter fields on the brane that satisfy these conditions, one can hence ensure 
that the geometry will be as depicted on the right hand side of figure 
\ref{fig::strip}.

%%%%%%%%%%%%%%%%%%%%%%%%%%%%%%%%%%%%%%%%%%%%%%%%%%%%%%%%%%%%%%%%%%%%%%%%%%%%%%%%
%%%%%%%%%%%%%%%%%%%%%%%%%%%%%%%%%%%%%%%%%%%%%%%%%%%%%%%%%%%%%%%%%%%%%%%%%%%%%%%%

\section{Geodesic normal flows}
\label{app::normalflow}

Here we investigate the question if and under which
conditions solutions to the Israel junction conditions 
\eqref{eq:Israeljunctionconditions} with constant 
brane tension as in section \ref{sec::toymodel} and
\ref{sec::Kondo} can be obtained by a normal flow starting from some 
non-backreacted solution. 
The governing equations of motion are assumed to 
be given by either imposing von Neumann conditions, as in the study of AdS/BCFT
(see section \ref{sec::AdSBCFTreview} and references therein) or by the Israel
junction conditions, as in the study of AdS/DCFT (see section 
\ref{sec::backreaction}).

In both cases, as discussed in sections \ref{sec::AdSBCFTreview} and 
\ref{sec::backreaction}, the equations of motion for the embedding are given by
\eqref{eq:leftoverIsrael}
\begin{equation}
 K_{ij} = -\frac{\kappa \lambda}{2} \left(\frac{1}{d-1}\right) \, \gamma_{ij}
 \label{eq:IJCproportionality}
\end{equation}
where the proportionality constant is 
fixed by the dimension of the embedding $d$, the gravitational coupling 
constant $\kappa$ and the constant brane tension $\lambda$.
Here, both $K_{ij}$ and $\gamma_{ij}$ are (0,2)-tensors on the embedded 
hypersurface.

As was discussed in the main text, starting from an initially totally geodesic
embedding, which solves the Israel junction conditions under the above 
assumptions for $\kappa=0$, we can generate solutions for constant brane 
tensions in AdS (see section \ref{sec::AdSBCFTreview} and 
\cite{Azeyanagi:2007qj}) and BTZ backgrounds (see section \ref{sec::Kondo}) by
following geodesic normal flows.

How does this approach generalise to other spacetimes and which assumptions have 
to be satisfied?
We show that the construction continues to work as long as the ambient 
geometry is given by an Einstein manifold and the proportionality of the induced
metric and the extrinsic curvature holds at the initial surface from where we 
start to follow the flow. 
This is always the case if we start from a
totally geodesic embedding in any Einstein manifold, which resembles the probe 
limit.
We furthermore derive that the flow acts on the induced metric as a conformal 
transformation.

Suppose we have an embedding $X_0: Q \hookrightarrow \mathcal{N}$ of some 
$(D-1)$-dimensional manifold $Q$ in a $D$-dimensional manifold 
$\mathcal{N}$.\footnote{
In the main text, we denote $\mathcal{N}$ as $N^+$ but since we will refer to 
the normal vector field as $N$, we changed notation.} We define a family of 
embeddings $X_s: Q \hookrightarrow \mathcal{N}$ by following the 
geodesic normal flow starting from the initial embedding $X_0$ for an arc length
$s$.
The vector field associated to the flow is denoted by $N$. 
It is normalised and satisfies the geodesic equation by construction.
Gauss' lemma states that this construction yields a family of 
regular embeddings at least in the vicinity of the initial surface and that the 
vector field generating these flows is always orthogonal to every member of the 
family of embeddings.

To show whether \eqref{eq:IJCproportionality} holds everywhere along the flow,
we need to derive how the induced metric and the extrinsic curvature of the 
family of embeddings change subject to the geodesic normal flow.
For this purpose, it is beneficial to define these geometric quantities not as 
tensors on the codimension one embeddings, but rather as tensors in the ambient
spacetime.

The induced metric $\gamma$ of the family of embeddings is defined by
\begin{equation}
 \gamma = g - n \otimes n, \qquad \gamma_{\mu\nu} = g_{\mu\nu} - N_{\mu}N_{\nu},
\end{equation}
where $n = g(N,\cdot) = N_{\alpha} \totd x^{\alpha}$ 
denotes the associated 1-form to the normal vector field $N$. This is a 
(0,2)-tensor in the ambient spacetime defined everywhere along 
the flow and projects the ambient metric $g$ onto the embeddings in the sense 
that $\gamma(N,\cdot)=0$.
Its change along the flow is given by the Lie derivative
\begin{equation}
 \mL_N \gamma 
 = \mL_N g - \mL_n (n \otimes n) 
 = \mL_N g - \mL_N(n) \otimes n - n \otimes \mL_n(n)
 = \mL_N g,
\end{equation}
where we used $\mL_N(n)=0$ if $N$ generates a geodesic flow. Hence, we find
\begin{equation}
 (\mL_N \gamma)_{\mu\nu} 
 = (\mL_N g)_{\mu\nu} 
 = \nabla_{\mu}N_{\nu}+\nabla_{\nu}N_{\mu}.
\end{equation}

We define the extrinsic curvature $K$ by
\begin{equation}
 K(U,V) := g(N,\nabla_U V) = -g(\nabla_U N,V).
\end{equation}
In components, this reads
\begin{equation}
 K_{\mu\nu} 
 = -\nabla_{\mu} N_{\nu} = -\nabla_{\nu} N_{\mu} = -\nabla_{(\mu} N_{\nu)}
 =-\frac{1}{2} \left(\nabla_{\mu}N_{\nu} + \nabla_{\nu}N_{\mu}\right)
 \label{eq:extrCurvCoords}
\end{equation}
from which we see that it satisfies $K(N,\cdot)=0$.
Summarising the above, we obtained that 
\begin{equation}
(\mL_N \gamma)_{\mu\nu} = - 2 K_{\mu\nu},
\end{equation}
which is a standard result in differential geometry and may be seen as
an alternative definition for the extrinsic curvature.

Next, let us compute the Lie derivative of the extrinsic curvature $K$. 
For $N$ generating a geodesic flow, its Lie derivative is given by
\begin{equation}
\begin{aligned}
 (\mathcal{L}_NK)_{\mu\nu} 
 =& -\frac{R}{(D-2)(D-1)} \gamma_{\mu\nu}
    + N^{\alpha}N^{\beta} C_{\alpha\mu\beta\nu}
    - g^{\alpha\beta} K_{\alpha\mu}\,K_{\beta\nu} \\
  & + \frac{1}{D-2}\left(
    N^{\alpha}N^{\beta}R_{\alpha\beta}g_{\mu\nu} 
    + R_{\mu\nu}
    -N^{\alpha}R_{\alpha\nu}N_{\mu}
    -N^{\beta}R_{\beta\mu}N_{\nu}\right),
\end{aligned}
\label{eq:LieDerivativeExtrinsicCurvature}
\end{equation}
with $C$ the Weyl curvature tensor.
For our purposes, it is interesting to look at this equation for more specific 
ambient manifolds, especially for Einstein manifolds for which 
\begin{equation}
 R_{\mu\nu} = \frac{R}{D}\, g_{\mu\nu} \,.
\end{equation}
Let us further assume that the manifolds have vanishing Weyl curvature, $C=0$, 
and hence are conformally flat.
For such manifolds, \eqref{eq:LieDerivativeExtrinsicCurvature} simplifies 
drastically to
\begin{equation}
 (\mathcal{L}_NK)_{\mu\nu} = 
 \frac{R}{D(D-1)}\gamma_{\mu\nu} 
 - g^{\alpha\beta} K_{\mu\alpha}\,K_{\nu\beta}\,.
 \label{eq:LieDerivativeExtrinsicCurvatureEinstein}
\end{equation}

In an especially appropriate coordinate system in which the normal is given
by the unit vector $N^{\alpha}=\{1,0,\ldots\}$, the equations are given by
\begin{align}
 \partial_s \gamma_{\mu\nu} &= - 2 K_{\mu\nu} \,,\\
 \partial_s K_{\mu\nu} &= \tilde{R} \,\gamma_{\mu\nu} -
 g^{\alpha\beta}K_{\mu\alpha}K_{\nu\beta},
\end{align}
where we defined $\tilde{R} =R/(D(D-1))$.
This is a coupled system of first order ODEs in $s$ and hence admits a unique 
solution depending only on the initial conditions $\gamma(0)$ and $K(0)$.

We now \emph{assume} a particular ansatz for the solution of the form
\begin{equation}
 K_{\mu\nu}(s) = c(s) \gamma_{\mu\nu}(s)
\end{equation}
which we require to be satisfied at $s=0$.
Deriving w.r.t.~$s$ and comparing with the evolution equations of the flow, we
obtain an ordinary differential equation for $c(s)$ of the form
\begin{equation}
 c'(s)=\tilde{R}+c^2(s),
\end{equation}
which can solved by separation of varibles with the unique solution
\begin{equation}
 c(s) =
 \begin{cases} 
  -\sqrt{|\tilde{R}|}\,\tanh\left(\sqrt{|\tilde{R}|}\,s+s_0\right) 
  &\text{if} \,\,\tilde{R}<0  
  \\
  (s_0-s)^{-1}
  &\text{if} \,\,\tilde{R}=0
  \\
  \sqrt{\tilde{R}}\,\tan\left(\sqrt{\tilde{R}}\,s + s_0\right) 
  &\text{if} \,\,\tilde{R}>0
 \end{cases}\,.
\end{equation}

We find two parameters in this set of solutions. One is given by $\tilde{R}$ 
which describes the ambient geometry. The other parameter, $s_0$, defines the
proportionality constant initially, at $s=0$. It makes sense intuitively that
the flow must at least include those two parameters.

There are three cases to be considered for an initially totally geodesic 
embedding with $c(0) = 0$, for which only the scalar curvature $R$ governs the 
sign of the extrinsic curvature: 

\paragraph{Negative scalar curvature} is the case important for our 
considerations in holography
where the ambient spacetime is given by locally AdS in 2+1 dimensions with 
$\Lambda<0$. 
In particular in 2+1 dimensions, the results are valid for Poincar\'e AdS, 
global AdS, thermal AdS and BTZ black holes\footnote{In this work we are 
interested in the case of a junction along a brane with timelike worldvolume. 
Some interesting constructions involving BTZ black holes and junctions along 
spacelike hypersurfaces with constant tension were investigated in 
\cite{Loran:2010qn,Loran:2010zy}.}.
For $\lambda>0$, the backreacted geometry will be given by a positive value of 
$s$ and hence the volume increases for larger tensions $\lambda$. Furthermore, 
the proportionality constant $c(s)$ is bounded from above and below by
\begin{equation}
 |c(s)| < \sqrt{\tilde{R}}
\end{equation}
and comparing to \eqref{eq:IJCproportionality} exactly reproduces the bounds on 
the matter content found in sections \ref{sec::toymodel} and \ref{sec::Kondo}.

\paragraph{Vanishing scalar curvature} is given by an ambient Minkowski 
spacetime. If $c(0)=0$ initially, we find that $c(s)=0$ is the exact solution,
so the extrinsic curvature is not driven away from zero if the embedding
is maximally geodesic initially. Hence the construction does not work in flat 
space.

\paragraph{Positive scalar curvature} corresponds e.g.~to de Sitter space. 
In this case we must follow the geodesic normal flow to negative values of $s$ 
and hence the volume of the spacetime will be reduced in the backreacted 
geometry. It is remarkable that if we start with a maximally geodesic embedding 
($s_0=0$), the proportionality constant blows up at $|s|=\pi /(2\tilde{R})$.

This concludes our study of geodesic normal flows. We have shown that constant
brane tension solutions can be constructed by following such a flow starting
from an initially totally geodesic embedding with $K_{\mu\nu}=0$, if we assume
the ambient manifold to be Einstein with vanishing Weyl curvature.

%=========================================================================
%======================         END OF Tex       =========================
%=========================================================================

%\bibliography{refs}{}
%\bibliographystyle{JHEP}

\providecommand{\href}[2]{#2}\begingroup\raggedright\endgroup

%\printindex
 
\end{document}